\documentclass[a4paper,table]{article}

\usepackage[margin=1in]{geometry} 

\usepackage{graphicx} 
\usepackage{mathtools}
\usepackage{amsmath}
\usepackage{amsfonts}
\usepackage{amssymb}
\usepackage{amsthm}
\usepackage{enumitem}
\setlist[enumerate]{leftmargin=*}
\usepackage{subcaption}
\usepackage{tikz}
\usepackage{tabularx}
\usepackage{quiver}
\usepackage{todonotes}
\usepackage{hyperref}
\usepackage{cleveref} 
\usepackage[normalem]{ulem}
\usepackage{authblk}

\definecolor{qubitcolor}{rgb}{1,1,1}
\definecolor{xcheckcolor}{rgb}{0.89,0,0.13}
\definecolor{zcheckcolor}{rgb}{0,0.5,1}
\newcommand{\dhor}{\mathsf{h}}
\newcommand{\dver}{\mathsf{v}}
\newcommand{\hor}{\mathsf{v}}
\newcommand{\ver}{\mathsf{h}}
\newcommand{\op}[1]{\operatorname{#1}}
\newcommand{\lmp}[1]{\operatorname{L}_{#1}}
\newcommand{\rmp}[1]{\operatorname{R}_{#1}}
\newcommand{\im}{\operatorname{im}}
\newcommand{\ann}[1]{\operatorname{ann}(#1)}
\newcommand{\Code}{\mathcal{C}}
\newcommand{\ftwo}{\mathbb{F}_2}
\DeclareMathOperator{\id}{id}
\DeclareMathOperator{\Tor}{Tor}

\newtheorem{theorem}{Theorem}[section]

\newtheorem{proposition}[theorem]{Proposition}

\newtheorem{corollary}[theorem]{Corollary}

\theoremstyle{definition}
\newtheorem{definition}[theorem]{Definition}
\newtheorem{example}[theorem]{Example}
\newtheorem{remark}[theorem]{Remark}
\crefname{proposition}{Proposition}{Propositions}

\title{Logical Operators and Fold-Transversal Gates of \\Bivariate Bicycle Codes}

    \author{Jens Niklas Eberhardt$^1$\thanks{mail{@}jenseberhardt.com} \hspace{2cm} Vincent Steffan$^2$\thanks{vincent.steffan{@}meetiqm.com}}
\date{
	$^1$Mathematical Institute of the University of Bonn, Germany   
 \\%
	 $^2$IQM Quantum Computers, Germany \\\vspace{2em} 
	\today
}

\begin{document}

\maketitle
\begin{abstract}
Quantum low-density parity-check (qLDPC) codes offer a promising route to scalable fault-tolerant quantum computation with constant overhead. Recent advancements have shown that qLDPC codes can outperform the quantum memory capability of surface codes even with near-term hardware. The question of how to implement logical gates fault-tolerantly for these codes is still open. We present new examples of high-rate bivariate bicycle (BB) codes with enhanced symmetry properties. These codes feature explicit nice bases of logical operators (similar to toric codes) and support fold-transversal Clifford gates without overhead. As examples, we construct 
$[[98,6,12]]$ and $[[162, 8, 12]]$ BB codes which admit interesting fault-tolerant Clifford gates. Our work also lays the mathematical foundations for explicit bases of logical operators and fold-transversal gates in quantum two-block and group algebra codes, which might be of independent interest. 
\end{abstract}

\section{Introduction}
\subsection{Motivation}
Quantum low-density parity-check (qLDPC) codes yield a prom\-ising pathway to scalable fault-tolerant quantum computation with constant overhead \cite{gottesmanFaulttolerantQuantumComputation2014}. We refer to \cite{breuckmannQuantumLowDensityParityCheck2021a} for an overview of recent developments. In particular, a construction via products of classical codes devised in \cite{breuckmannBalancedProductQuantum2021b} yields qLDPC codes that were shown to be asymptotically good in \cite{panteleevAsymptoticallyGoodQuantum2022a}. 

Recently, \cite{bravyi2023highthreshold} showed that qLDPC codes allow practical fault-tolerant memories that outperform surface codes -- even for near-term quantum computers. In particular, they construct examples of high-rate \emph{bivariate bicycle (BB) codes}, see \cite{PhysRevA.88.012311}, and demonstrate their practicability via low-depth syndrome measurement and decoding algorithms.

Here, we devise new examples of BB codes with comparable parameters but with extra symmetry properties. We show that these codes have nice \emph{explicit bases of logical operators} and admit certain \emph{fold-transversal Clifford gates}, see \cite{breuckmannFoldTransversalCliffordGates2024}, with no overhead.
For example, we consider a $[[98, 6, 12]]$ BB code and construct fold-transversal gates generating the group $C_2\times\op{Sp}_2(\mathbb{F}_{2^3})$ (modulo Pauli operators). 
We achieve this by laying the mathematical foundations for explicit bases of logical operators and fold-transversal gates in quantum two-block (group algebra) codes, which may be of independent interest.

\subsection{Logical operators of BB codes} To get a good grip on logical gates of a quantum code, it is necessary first to gain a precise understanding of its logical operators. Computing just \emph{some} basis of logical operators is simple -- it relies on solving some linear equations. However, a basis computed this way does in general not give any insights into the structure of the code. In particular, for BB codes it does not reflect their beautiful symmetrical structure and is not adapted to the logical gates we will consider here.

BB codes are a generalization of \emph{hypergraph products} (HGP) of \emph{cyclic codes} (and a special case of \emph{balanced product codes}). For HGP codes, the \emph{Künneth formula} yields a basis of logical operators derived from the code words of the underlying classical codes, see \cite{Tillich_2014}. For example, the toric code is a HGP of two repetition codes and its vertical and horizontal logical operators arise from the non-trivial code word of each of the two repetition codes.

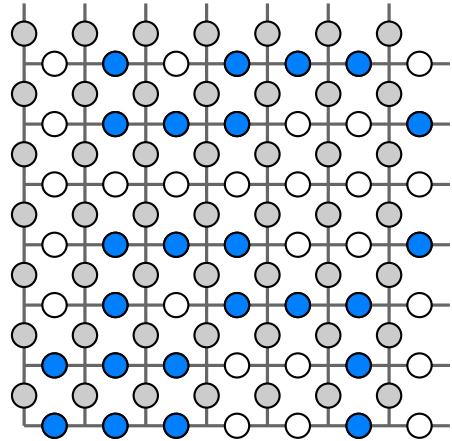
\begin{figure}
\begin{subfigure}[b]{0.45\textwidth}
    \centering
\begin{tikzpicture}[scale = .8]

\pgfmathsetmacro{\zstabxposition}{2}
\pgfmathsetmacro{\zstabyposition}{0}

\pgfmathsetmacro{\xstabxposition}{7}
\pgfmathsetmacro{\xstabyposition}{5}

    \draw[fill = zcheckcolor] (\zstabxposition,\zstabyposition) rectangle ++(1,1);

        \foreach \x in {1,2,3,4,5,6,7} (
            \draw[very thick, color = black!60!white] (\x,-1) -- (\x, 6);
        )
        \foreach \x in {-1,0,1,2,3,4,5} (
            \draw[very thick, color = black!60!white] (1,\x) -- (8,\x);
        )

        \foreach \x in {1,2,3,4,5,6,7} (
            \foreach \y in {-1,0,1,2,3,4,5} (
                \draw[thick, fill = white] (\x + .5, \y) circle (0.2);
            )
        )
        \foreach \x in {1,2,3,4,5,6,7} (
            \foreach \y in {0,1,2,3,4,5,6} (
                \draw[thick, fill = black!20!white] (\x , \y - .5) circle (0.2);
            )
        )



    \draw[thick, fill = zcheckcolor] (\zstabxposition+1,\zstabyposition + .5) circle (0.2);
    \draw[thick, fill = zcheckcolor] (\zstabxposition + 0,\zstabyposition + 3.5) circle (0.2);
    \draw[thick, fill = zcheckcolor] (\zstabxposition + 0,\zstabyposition + 4.5) circle (0.2);
    \draw[thick, fill = zcheckcolor] (\zstabxposition + .5,\zstabyposition + 1) circle (0.2);
    \draw[thick, fill = zcheckcolor] (\zstabxposition + 3.5,\zstabyposition ) circle (0.2);
    \draw[thick, fill = zcheckcolor] (\zstabxposition +4.5,\zstabyposition ) circle (0.2);

    \draw[fill = xcheckcolor] (\xstabxposition - .2, \xstabyposition - .2) rectangle ++(.4,.4);

    \draw[thick, fill = xcheckcolor] (\xstabxposition  - .5, \xstabyposition ) circle (0.2);
    \draw[thick, fill = xcheckcolor] (\xstabxposition  + .5, \xstabyposition -3) circle (0.2);
    \draw[thick, fill = xcheckcolor] (\xstabxposition  + .5, \xstabyposition -4) circle (0.2);

    \draw[thick, fill = xcheckcolor] (\xstabxposition  , \xstabyposition -.5) circle (0.2);
    \draw[thick, fill = xcheckcolor] (\xstabxposition  -3, \xstabyposition +.5) circle (0.2);
    \draw[thick, fill = xcheckcolor] (\xstabxposition  -4, \xstabyposition +.5) circle (0.2);
 \end{tikzpicture} 
 \caption{The support of an $X$- (red) and $Z$-stabilizer (blue) associated to a vertex and a face of the grid. There are $49$ such $X$- and $Z$-stabilizers, each.}\label{subfig:stabilizers}
 \end{subfigure}
 \hfill
 \begin{subfigure}[b]{0.45\textwidth}
    \centering
\begin{tikzpicture}[scale = .8]

        \foreach \x in {1,2,3,4,5,6,7} (
            \draw[very thick, color = black!60!white] (\x,-1) -- (\x, 6);
        )
        \foreach \x in {-1,0,1,2,3,4,5} (
            \draw[very thick, color = black!60!white] (1,\x) -- (8,\x);
        )

        \foreach \x in {1,2,3,4,5,6,7} (
            \foreach \y in {-1,0,1,2,3,4,5} (
                \draw[thick, fill = white] (\x + .5, \y) circle (0.2);
            )
        )
        \foreach \x in {1,2,3,4,5,6,7} (
            \foreach \y in {0,1,2,3,4,5,6} (
                \draw[thick, fill = black!20!white] (\x , \y - .5) circle (0.2);
            )
        )

 \foreach \x/\y in {1/3,1/2,2/0,1/1,1/0,2/5,1/6,3/3,1/5,3/2,2/3,4/6,2/1,3/6,3/5,5/2,5/1,4/2,5/0,6/5,5/6,6/3,0/1,0/0} (
 \draw[thick, fill = zcheckcolor] (\x+ 1.5,\y-1) circle (.2); 
 )

 \end{tikzpicture} 
 \caption{A logical $Z$-operator, supported on horizontal edges. It generates all other logical $Z$-operators via translations and reflection along the diagonal.}\label{subfig: logical operator}
 \end{subfigure}
    \caption{A $[[98, 8, 12]]$ BB code $\Code(c,d)$ for $c= x + y^3 + y^4$ and $d= y + x^3 + x^4$. Physical qubits correspond to vertical and horizontal edges and stabilizers to vertices and faces on a $7\times 7$-grid.} \label{fig:77code}
\end{figure}

The Künneth formula is however \emph{wrong} for general BB codes. To address this, we introduce a notion of \emph{purity}, which we explain now. Recall that 
a BB code, say $\Code(c,d)$, is defined by two bivariate polyomials $c(x,y), d(x,y)$ and non-negative integers $\ell,m.$ The physical qubits of a BB code can be identified with the horizontal and vertical edges of an $\ell\times m$ square lattice, see \Cref{fig:77code}. 
A BB code is \emph{pure}, if it admits a basis of logical operators which are supported on either vertical or horizontal edges only.
A logical operator supported purely on vertical (or horizontal) edges comes from a code word of the classical \emph{2D-cyclic code}, see \cite{IMAI19771}, with check polynomial $c(x,y)$ (or $d(x,y)$).
This brings us back to the world of classical codes. We show various, easy-to-check conditions on when BB codes are pure. For example, we show that all BB codes are pure if $\ell$ and $m$ are odd.
We note that certain pure logical operators are also discussed in \cite[Section 9.1]{bravyi2023highthreshold}.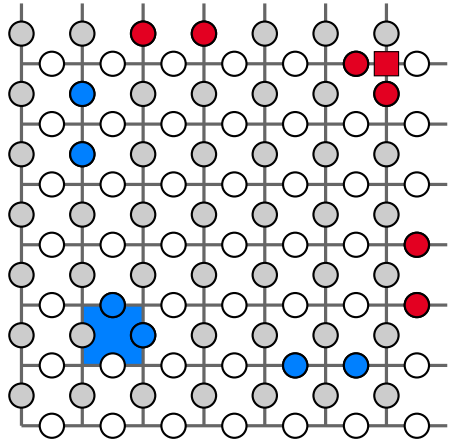

Another favorable property of HGPs of cyclic codes is that all code words of the underlying cyclic codes are generated by translates of a single code word (the generator polynomial). Again, this is \emph{wrong} for BB codes, since the underlying 2D-cyclic codes do not have this property in general. To address this, we introduce the notion of \emph{principal} BB codes. These are pure BB codes whose logical operators are generated by the translates of a single vertical and a single horizontal logical operator. 
We show that if $\ell$ and $m$ are odd, all BB codes are principal. Moreover, in \Cref{sec:semiperiodic} we discuss so-called \emph{semiperiodic} 2D-cyclic codes, which have particularly nice generators. We list some BB codes with these properties and favorable parameters in \Cref{tab:properties}, and refer to \Cref{fig:77code} and \Cref{sec:examples} for explicit examples. 
\begin{table}[h]
\centering
\definecolor{highlightcolor}{rgb}{0.8,0.9,0.7} 
\definecolor{symhighlightcolor}{rgb}{0.7,0.8,0.5} 

\newcommand{\highlightrow}{\rowcolor{highlightcolor}}
\newcommand{\symhighlightrow}{\rowcolor{symhighlightcolor}}
\begin{tabular}{|c|c|c|c|c|c|c|}
\hline
$c$ & $d$ & $l, m$ & $[[n, k, d]]$ & Pure & Prin. & Sym. \\
\hline
$1 + y + y^5$ & $y^3 + x + x^2$ & $3, 15$ & $[[90, 8, 10]]$ & $\checkmark$ & $\checkmark$ & $\times$ \\
$x^3 + y + y^2$ & $y^3 + x + x^2$ & $6, 12$ & $[[144, 12, 12]]$ & $\times$ & $\times$ & $\times$ \\
\hline
\highlightrow
$1 + y + y^2$ & $y^3 +  x^2+ x^4 $ & $6, 9$ & $[[108, 16, 6]]$ & $\checkmark$ & $\checkmark$ & $\times$ \\ 
\highlightrow
$ x^2 + y+ y^3+y^4  $ & $y^2 +  x+ x^3+x^4  $ & $8, 8$ & $[[128, 14, 12]]$ & $\times$ & $\times$ & $\checkmark$ \\ 
\highlightrow
$1 + x + y$ & $x^3+ y + y^2 $ & $9, 9$ & $[[162, 4, 16]]$ & $\checkmark$ & $\checkmark$ & $\times$ \\
\highlightrow
$1+x + y^6 $ & $y^3+ x^2 + x^3 $ & $9, 9$ & $[[162, 12, 8]]$ & $\checkmark$ & $\checkmark$ & $\times$ \\
\highlightrow
$1+ y + y^2 $ & $ y^3 + x^3 + x^6$ & $9, 9$ & $[[162, 24, 6]]$ & $\checkmark$ & $\checkmark$ & $\times$ \\
\highlightrow
$x^3 + y + y^2$ & $y^3 + x + x^2$ & $9, 15$ & $[[270, 8, 18]]$ & $\checkmark$ & $\checkmark$ & $\times$ \\
\hline
\symhighlightrow
$x+ y^3 + y^4$ & $y+x^3 + x^4$ & $7, 7$ & $[[98, 6, 12]]$ & $\checkmark$ & $\checkmark$ & $\checkmark$ \\
\symhighlightrow
$x^3 + y + y^2$ & $y^3 + x + x^2$ & $9, 9$ & $[[162, 8, 12]]$ & $\checkmark$ & $\checkmark$ & $\checkmark$ \\
\hline
\end{tabular}

\caption{BB codes $\Code(c,d)$ on an $\ell\times m$-lattice and their parameters. For the first two, see \cite{bravyi2023highthreshold}. We found the codes highlighted in green by a computer search. The last two codes are discussed in detail in \Cref{sec:examples}. The last columns denote if the codes are \emph{pure} (see \Cref{def:pure}), \emph{principal} (see \Cref{def:principal}) and \emph{symmetric} (see \Cref{def:symmetric}). These are favorable properties, inspired by the toric code, which allow us to construct logicals and gates in a natural way. For the last two codes, we construct fold-transversal gate groups isomorphic to $C_2\times \op{Sp}_{2}(\mathbb{F}_{2^3})$ and $\op{Sp}_2(\mathbb{F}_{2^2})\times (\op{Sp}_2(\mathbb{F}_{2^2})\rtimes C_2)$, respectively, see \Cref{sec:examples}}

\label{tab:properties}
\end{table} 

Our definitions and results for BB codes apply more generally to quantum codes constructed from commuting operators (so-called \emph{quantum two-block codes}, see \cite{linQuantumTwoblockGroup2023,wangAbelianNonabelianQuantum2023a}) and we discuss the homological and commutative algebra of such codes in \Cref{sec:homologicalalgebraofcodesfromcummutingoperators}. The adventurous reader may find a conceptual explanation of these results in terms of the \emph{Künneth spectral sequence} for balanced product codes in \Cref{sec:balancedproductkuenneth}. Related to this, we also explain how to obtain higher-dimensional generalizations of BB codes via repeated balanced products in \Cref{sec:nblockcodes}. This might be interesting in the context of \emph{single-shot} error correction, see \cite{Bomb_n_2015}, and achieving fold-transversal gates in higher levels of the Clifford-hierarchy.

\subsection{Fold-transversal gates}
For the toric and surface codes, Clifford gates can be implemented without overhead \cite{moussaTransversalCliffordGates2016a}, by `folding the code along the diagonal'. This has been generalized to CSS codes in \cite{breuckmannFoldTransversalCliffordGates2024} where the notion of \emph{fold-transversal gates} is introduced. These are certain transversal gates associated with \textit{automorphisms} and $Z\!X$\textit{-dualities} of the quantum code. A $Z\!X$-duality is an isomorphism between the code and exchanges $X$- and $Z$-checks. {In \Cref{sec:groupalgebracodes}, we discuss $Z\!X$-dualities and fold-transversal gates in the broader context of quantum codes derived from group algebras (so-called \emph{quantum two-block group algebras codes}, see \cite{linQuantumTwoblockGroup2023}).}

We then apply this framework to BB codes in \Cref{sec:bblogicals}. We obtain the following fold-transversal gates. Shifting the lattice in the $x$ and $y$ direction yields automorphisms of the code and so-called \emph{swap-type gates}
$\op{Swap}_x,\op{Swap}_y.$ These gates have a particularly nice explicit description related to our choice of basis of logical operators. Exchanging horizontal and vertical edges yields a $Z\!X$-duality $\tau_0$ and a so-called \emph{Hadamard-type gate} $H_{\tau_0}$.
We are particularly interested in \emph{symmetric} BB codes for which $c(x,y)=d(y,x)$ and $\ell=m.$ In this case, we obtain an additional \emph{CZ-type gate} $\op{CZ}_{\tau_0\omega}$ which arises from a $Z\!X$-duality exchanging $x$ and $y$ and flipping the axes. {We describe these gates and the group generated by them in detail for two explicit examples in \Cref{sec:examples}. We note that \cite[Section 9.3 and 9.4]{bravyi2023highthreshold} also discusses fold-transversal gates for the BB codes. There, certain swap-type gates and a circuit design for a particular $ZX$-duality are provided.}

\subsection{Outline} In Section~\ref{sec:homologicalalgebraofcodesfromcummutingoperators}, we briefly recall the connection between homological algebra and CSS codes. We then introduce the central concepts of \textit{pure} and \textit{principal} codes and prove several characterizations for pure and principal codes. We then recall the definition of group algebra codes~\cite{linQuantumTwoblockGroup2023} and the theory of and $Z\!X$-dualities and fold-transversal gates~\cite{breuckmannFoldTransversalCliffordGates2024} in~\Cref{sec:groupalgebracodes} and discuss how fold-transversal gates arise in the context of group algebra codes. In~\Cref{sec:BBcodes}, we refine this discussion to the special case of bivariate bicycle codes. Finally, in \Cref{sec:examples}, we showcase how these results unfold for two particular examples.

\subsection{Acknowledgments} We thank Nikolas Breuckmann, Martin Leib, Pedro Parrado, and Francisco Pereira for many helpful discussions.
\section{Homological algebra of CSS codes of commuting operators}\label{sec:homologicalalgebraofcodesfromcummutingoperators}
We first briefly recall CSS codes and their relation to chain complexes. This homological perspective on CSS codes allows us to discuss a general mathematical framework to understand the logical operators in various families of quantum codes, such as \emph{bivariate bicycle codes} \cite{PhysRevA.88.012311} and \emph{quantum two-block codes} \cite{wangAbelianNonabelianQuantum2023a}. The reader only interested in particular examples may skip this section.
\subsection{CSS codes and chain complexes}
A quantum CSS code $\mathcal{Q}\subset (\mathbb{C}^2)^{\otimes n}$ is defined by two matrices $H_X \in \mathbb{F}_2^{r_X \times n}$ and $H_Z \in \mathbb{F}_2^{r_Z \times n}$ with $H_XH_Z^{\op{tr}}=0.$
Each row in $H_X$ and $H_Z$ determines a Pauli-X and Pauli-Z operator, respectively, which act on the qubits in the support of the row. The condition $H_XH_Z^{\op{tr}}=0$ ensures that all Pauli-X and Pauli-Z operators have even overlap and hence commute. The subspace $\mathcal{Q}$ is the joint $(+1)$-eigenspace of these operators.

Matrices $H_X$ and $H_Z$ with $H_XH_Z^{\op{tr}}=0$ arise naturally in homological algebra.
A chain complex $\mathcal{C}$ with three terms
\begin{equation*}
\begin{tikzcd}[row sep=small]
   \mathcal{C}\colon C_0 & C_1 & C_2
    \arrow["d_1"', from=1-2, to=1-1]
    \arrow["d_2"', from=1-3, to=1-2]
\end{tikzcd}
\end{equation*}
consists of two linear maps $d_1$ and $d_2$, called differentials, such that $d_1d_2=0.$ If the terms $C_i$ are $\ftwo$-vector spaces equipped with finite ordered bases, one may take $H_X$ and $H_Z^{\op{tr}}$ to be the matrices associated to the differentials $d_1$ and $d_2.$

In this language, logical $Z$-operators of the code $\mathcal{Q}$ correspond to the first homology group $H_1(\mathcal{C})=\ker(d_1)/\im(d_2)$ of the complex. The logical $X$-operators arise from the dual space $H_1(\mathcal{C})^*=H^1(\mathcal{C}).$
In the following, we will often identify a CSS code with the defining chain complex.
\begin{remark}
    In the following, we will focus on logical $Z$-operators and hence on the homology of the complex $H_1(\mathcal{C}).$ By passing from the complex $\mathcal{C}$ to its dual $\mathcal{C}^*$, this translates to logical $X$-operators, see also \Cref{rmk:zxtopassfromztox}.
\end{remark}
\subsection{Quantum codes from commutative rings}\label{sec:quantumcodesfromcommutativerings}
Let $R$ be a commutative ring\footnote{This setup naturally generalizes to bimodules over rings that are not necessarily commutative, see \Cref{app:bimodules} and \cite{linQuantumTwoblockGroup2023,wangAbelianNonabelianQuantum2023a}} and $c,d\in R.$ Then, multiplication with $c$ and $d,$ respectively, provides a commutative diagram
\begin{equation}\label{eq:doublecomplexofCcd}
\begin{tikzcd}
	R_{\dver} & R \\
	R & R_{\dhor}.
	\arrow["d"', from=1-1, to=2-1]
	\arrow["c"', from=1-2, to=1-1]
	\arrow["d", from=1-2, to=2-2]
	\arrow["c"', from=2-2, to=2-1]
\end{tikzcd}
\end{equation}
Here, we decorate two copies of $R$ as $R_{\dhor}$ and $R_{\dver}$ and call the corresponding elements \emph{horizontal} and \emph{vertical}. This naming convention is motivated by the toric code, see \Cref{ex:somebivariatecodes}.

By passing to the total complex, this diagram yields the following chain complex 
\begin{equation}
\label{eq:complexofCcd}
\begin{tikzcd}[row sep=small]
   \Code(c,d)\colon R & {R_{\dhor} \oplus R_{\dver}} & R
    \arrow["{(c, -d)}"', from=1-2, to=1-1]
    \arrow["\left(\begin{smallmatrix} d \\ c \end{smallmatrix}\right)"', from=1-3, to=1-2]
\end{tikzcd}
\end{equation}
that lives, from left to right, in degrees $0,1$ and $2$.
The differential squaring to $0$ corresponds to the fact that $c$ and $d$ commute. The first homology group of the complex is the subquotient of $R_{\dhor}\oplus R_{\dver}$ given by
$$
H=H(c,d)=H_1(\Code(c,d))=\frac{\{(f,g)\in R_{\dhor}\oplus R_{\dver} \mid cf=dg\}}{\{(rd,rc)\mid r\in R\}}
$$ and we denote homology classes by $[f,g]\in H$ for representatives $(f,g)\in R_{\dhor}\oplus R_{\dver}.$

Now assume that $R$ is a finite-dimensional commutative $\ftwo$-algebra equipped with an ordered basis. Then, all of the terms in the chain complex $\Code(c,d)$ are naturally equipped with an ordered basis as well. We hence obtain a CSS quantum code which we also denote by $\Code(c,d).$ The code is an example of a \emph{quantum two-block code}, see \cite{wangAbelianNonabelianQuantum2023a}.
The code has $2\dim R$ physical and $\dim H$ logical qubits.
\subsection{The fundamental exact sequence and pure classes}
For hypergraph product (HGP) quantum codes such as the toric code, the logical operators are generated by purely `horizontal' and `vertical' operators.
This is a consequence of the Künneth formula.

In this section, we study to what extent this is true for quantum codes derived from commutative rings as considered in \Cref{sec:quantumcodesfromcommutativerings}. The goal is to give a characterization of this property.
\begin{definition}\label{def:pure}
   Let $h\in H=H(\Code(c,d)).$ We call $h$ \emph{horizontally pure} if $h=[f,0]$ for $f\in R_{\dhor}$ and \emph{vertically pure} if $h=[0,g]$ for $g\in R_{\dver}.$ Moreover, $h$ is called \emph{pure} if it is either horizontally or vertically pure.
   We denote the subspaces generated by horizontally and vertically pure classes by $H_{\dhor},H_{\dver}\subset H$ and say that $H$ is \emph{pure} if $H=H_{\dhor}+ H_{\dver}.$
\end{definition}

Note that for $[f,0]\in H_{\dhor}$ we have $cf=0.$ Hence we obtain surjections
\begin{align*}
    \ann{c}&\to H_{\dhor}, f\mapsto [f,0] \text{ and }\\
    \ann{d}&\to H_{\dver}, g\mapsto [0,g]
\end{align*}
where for $m\in R$ we denote the annihilator by $\ann{m}=\{r\in R\mid rm=0\}.$
Pure homology classes can hence be constructed by determining $\ann{c}$ and $\ann{d}$ which brings us to the world of classical codes.

Purity is closely related to the `relative positions' of the ideals $(c)=cR$ and $(d)=dR$ in $R$ as well as the annihilators $\ann{c}$ and $\ann{d}$ as we will explain now.

Denote by
$$Z=\{(f,g)\in R_{\dhor}\oplus R_{\dver} \mid cf=dg\} \text{ and }B=\{(dr,cr)\mid r\in R\}
$$
so that $H=H(c,b)=Z/B.$ Our starting point is the short exact sequence describing $Z.$
\[\begin{tikzcd}[column sep=scriptsize,row sep=tiny]
	0 & {\ann{c}\oplus\ann{d}} & Z & {(c)\cap (d)} & 0 \\
	& {(f,g)} & {(f,g)} & {cf=dg.}
	\arrow[from=1-1, to=1-2]
	\arrow[from=1-2, to=1-3]
	\arrow[from=1-3, to=1-4]
	\arrow[from=1-4, to=1-5]
	\arrow[maps to, from=2-2, to=2-3]
	\arrow[maps to, from=2-3, to=2-4]
\end{tikzcd}\]
By passing to the `quotient by $B$' we obtain the following description of $H$.
\begin{theorem}\label{thm:fundamentalsequence}The following maps yield an exact sequence\footnote{This means that the kernel of each map coincides with the image of the previous map.}, which we call the \emph{fundamental exact sequence},
\begin{equation}\label{eq:fundamentalsequence}\begin{tikzcd}[column sep=scriptsize,row sep=tiny, column sep = 11pt]
	0 & {\frac{\ann{cd}}{M}} & {\frac{\ann{c}}{\ann{c}(d)} \oplus \frac{\ann{d}}{(c)\ann{d}}} & {H=Z/B} & {\frac{(c)\cap (d)}{(cd)}} & 0 \\
	& {[r]} & {([dr],[cr]),([f],[g])} & {[f,g]} & {[cf]=[dg]}
	\arrow[from=1-1, to=1-2]
	\arrow["\alpha", from=1-2, to=1-3]
	\arrow["\beta", from=1-3, to=1-4]
	\arrow["\gamma", from=1-4, to=1-5]
	\arrow[from=1-5, to=1-6]
	\arrow[maps to, from=2-2, to=2-3]
	\arrow[maps to, from=2-3, to=2-4]
	\arrow[maps to, from=2-4, to=2-5]
\end{tikzcd}\end{equation}
where $M$ is defined by $$M=\{r\in \ann{cd}\mid \exists f\in \ann{c}, g\in\ann{d}\text{ such that } rd=fd \text{ and } rc=gc\}.$$
Here we denote the respective equivalence classes in the quotient spaces by square brackets and the product of ideals $I,J\subset R$ by $IJ=\{\sum_i f_ig_i|f_i\in I, g_i\in J\}\subset R.$
The image of $\beta$ is the pure part $H_{\dhor}+H_{\dver}\subset H.$
\end{theorem}
\begin{proof}
One easily checks that $\alpha,\beta,\gamma$ are well-defined, $\beta\alpha=0$ and $\gamma\beta=0.$ We show that the sequence is exact in all four entries, starting from the right. 

First, let $r\in (c)\cap (d).$ Then $r=fc=gd$ for some $f,g\in R$ and hence
$$[r]=\gamma([f,g])\in \im(\gamma).$$

Second, let $[f,g]\in\ker(\gamma)$. Then $cf=dg=cdr$ for some $r\in R$. Then $(f-dr)\in \ann{c}$ and $(g-cr)\in \ann{d}$ and hence $$[f,g]=[f-dr,g+cr]=\beta([f-dr],[g-cr])\in \im{\beta}.$$ 
Third, let $([f],[g])\in \ker(\beta).$ Then $(f,g)=(rd,rc)$ for some $r\in R.$ Since $f\in \ann{c},$ we have $cdr=cf=0$ so that $r\in \ann{cd}.$ Hence $$([f],[g])=\alpha(r)\in \im(\alpha).$$
Fourth, $\alpha$ is injective by the definition of $M.$
\end{proof}
\begin{remark}
    The fundamental exact sequence is induced by the Künneth spectral sequence, see \Cref{sec:balancedproductkuenneth}.
\end{remark}
\begin{corollary}\label{cor:puritycriterion} The homology $H$ is pure, that is $H=H_{\dhor}+H_{\dver}$, if and only if $(c)\cap (d)=(cd).$ Moreover,  $H=H_{\dhor}\oplus H_{\dver}$ if and only if  $(c)\cap (d)=(cd)$ and $\ann{cd}=M.$ 
\end{corollary}
We also consider the following three interesting special cases.
\begin{corollary}
    If the ideals $(c)$ and $(d)$ are comaximal, that is $(c)+(d)=R$, then $H=0.$
\end{corollary}
\begin{proof} Assume that $(c)+(d)=R.$ Note that both $c$ and $d$ and hence $1\in R$ annihilate
$$\frac{(c)\cap (d)}{(cd)},{\frac{\ann{c}}{\ann{c}(d)}\; \text{ and } \;\frac{\ann{d}}{(c)\ann{d}}}.$$
A module annihilated by $1$ is zero. The statement follows from \Cref{thm:fundamentalsequence}.
\end{proof}
\begin{corollary}\label{cor:regular}
    If $R$ is a regular ring, then $H=H_{\dhor}+H_{\dver}$ implies $H=H_{\dhor}\oplus H_{\dver}.$
\end{corollary}
\begin{proof} By \Cref{cor:puritycriterion}, the statement is equivalent to showing that
$$\frac{(c)\cap (d)}{(cd)}=0\implies \frac{\ann{cd}}{M}=0.$$
By \Cref{sec:balancedproductkuenneth}, these two modules are isomorphic to $\Tor_1^R(R/(c),R/(d))$ and $\Tor_2^R(R/(c),R/(d)),$ respectively. Now by Tor-rigidity for regular rings, see \cite{lichtenbaumVanishingTorRegular1966}, 
$\Tor_1^R(R/(c),R/(d))=0$ implies that $\Tor_i^R(R/(c),R/(d))=0$ for all $i\geq 1.$
\end{proof}
\begin{corollary}\label{cor:prodoffieldspure}
    If $R$ is a product of fields, then $H=H_\dhor\oplus H_\dver.$
\end{corollary}
\begin{proof}
    In this case, $\op{Tor}_i(M,N)=0$ for all $i>0$ and all $R$-modules $M$ and $N$. Now we use \Cref{sec:balancedproductkuenneth}.
\end{proof}

The first term $\ann{cd}/M$ in the fundamental sequence also has a more direct but asymmetrical description, which can be useful.
\begin{proposition} \label{prop:descriptionoftor2}One of the conditions defining $M$ is vacuous and
\begin{align*}
    M&=\{r\in \ann{cd}\mid \exists f\in \ann{c}\text{ such that } rd=fd\}\\
    &=\{r\in \ann{cd}\mid \exists g\in\ann{d}\text{ such that }rc=gc\}.
\end{align*}
Moreover, the following maps are isomorphisms
\[\begin{tikzcd}[row sep=tiny]
	{\frac{(c)\cap\ann{d}}{(c)\ann{d}}} & {\frac{\ann{cd}}{M}} & {\frac{\ann{c}\cap (d)}{\ann{c}(d)}} \\
	{[rc]} & {[r]} & {[rd].}
	\arrow["{\mu_c}"', from=1-2, to=1-1]
	\arrow["{\mu_d}", from=1-2, to=1-3]
	\arrow[maps to, from=2-2, to=2-1]
	\arrow[maps to, from=2-2, to=2-3]
\end{tikzcd}\]
\end{proposition}
\begin{proof} Let $r\in \ann{cd}$ and $f\in \ann{c}$ such that $rd=fd.$ Let $g=r-f$ then $g\in \ann{d}$ and $gc=(r-f)c=rc.$ Hence $r\in M.$ This shows the first equality. The second equality follows in the same way.

We show that $\mu_c$ is an isomorphism. First, it is easy to see that $\mu_c$ is well-defined and injective using the description of $M$ above.
Now let $s=rc\in (c)\cap \ann{d}$ where $r\in R.$ Then $rcd=0$ so that $r\in \ann{cd}$ and $s=\mu_c(r).$ Hence $\mu_c$ is surjective and an isomorphism. That $\mu_d$ is an isomorphism follows in the same way.
\end{proof}
\subsection{Principal codes and relations of logical operators}
The fundamental sequence \eqref{eq:fundamentalsequence} shows that  $\Code(c,d)$ is closely related to the classical codes $\ann{c}$ and $\ann{d}.$ An interesting case is when these codes are generated by single elements.
\begin{definition}\label{def:principal}
    We call $\Code(c,d)$ \emph{principal} if it is pure and the ideals $\ann{c}$ and $\ann{d}$ are principal, that is, generated by single elements $P,Q\in R.$
\end{definition}
If the code is pure and principal, all logical operators can be generated from two code words. Moreover, $c$ and $d$ yield relations of the resulting code words.
\begin{corollary}
    Assume that $\Code(c,d)$ is principal. Write $\ann{c}=(P)$ and $\ann{d}=(Q)$. Then there is a surjection
    \begin{align}\label{eq:understandingallcodewords}
        R/(c,d)\oplus R/(c,d)\to H, ([f],[g])\mapsto [fP,gQ].
    \end{align}
    If, moreover, $\ann{P}=(c)$ and $\ann{Q}=(d),$ then the map is an isomorphism.
\end{corollary}
\begin{proof}
This follows from \Cref{thm:fundamentalsequence}.
\end{proof}
\begin{corollary} \label{cor:prodoffieldsiso}
    Assume that $R$ is a product of fields. Then $\Code(c,d)$ is principal, and the map in \eqref{eq:understandingallcodewords} is an isomorphism.
\end{corollary}
\begin{proof} A product of fields is a principal ideal ring and $\ann{\ann{m}}=(m)$ for all $m\in R.$
\end{proof}
\section{Group algebra codes and fold-transversal gates} \label{sec:groupalgebracodes}We now consider the special case of the codes in \Cref{sec:homologicalalgebraofcodesfromcummutingoperators} when $R$ is the group algebra of a finite commutative\footnote{The commutativity assumption can easily be circumvented when being careful about left and right multiplication. Since we are focused on commutative groups here, we will not do so.} group. The resulting codes are the \emph{two-block group-algebra (2BGA) codes} introduced in \cite{wangAbelianNonabelianQuantum2023a,linQuantumTwoblockGroup2023}.
We recall the explicit description of $X$- and $Z$-checks of the CSS-code $\Code(c,d)$ and describe how to construct fold-transversal gates using the framework of \cite{breuckmannFoldTransversalCliffordGates2024}.
\subsection{Group algebra codes} \label{sec:groupalgebracodedefinition}
Let $R=\ftwo[G]$ be the group algebra of a commutative finite group $G$ equipped with the basis $G.$ We fix $c,d\in R$ and $\op{supp}(c),\op{supp}(d)\subset G$ such that
$$c=\sum_{\mathclap{g\in \op{supp}(c)}} g \text{ and } d=\sum_{\mathclap{g\in \op{supp}(d)}} g.$$

Since $G$ is commutative, so is $R$, and we obtain a complex and quantum code $\Code(c,d)$ as defined in \Cref{sec:quantumcodesfromcommutativerings}.
In fact, the quantum code $\Code(c,d)$ is a so-called \emph{(two-block) group algebra code}, see \cite{wangAbelianNonabelianQuantum2023a}. 

We now explicitly describe the qubits and stabilizers.
The physical qubits of $\Code(c,d)$ can be partitioned in vertical and horizontal qubits indexed by $g_{\dhor}\in G_{\dhor}$ and $g_{\dver}\in G_{\dver}.$ The set of physical qubits $G_{\dhor}\cup G_{\dver}$ has $2|G|$ elements.
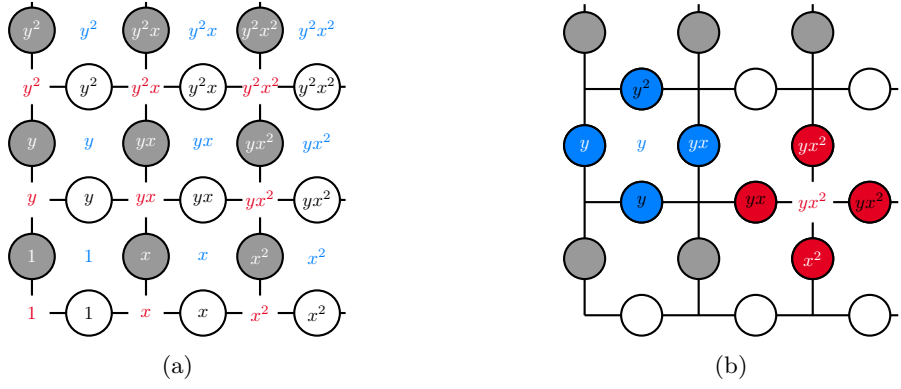
\begin{figure}[h]
\begin{subfigure}[b]{0.45\textwidth}
    \centering
\begin{tikzpicture}[scale = 1.5]
    \draw[thick] (0,0) -- (2.75,0);
    \draw[thick] (0,1) -- (2.75,1);
    \draw[thick] (0,2) -- (2.75,2);

    \draw[thick] (0,0) -- (0,2.75);
    \draw[thick] (1,0) -- (1,2.75);
    \draw[thick] (2,0) -- (2,2.75);

    \foreach \x in {0.5,1.5,2.5}(
    \foreach \y in {0,1,2}(
    \draw[fill = white,thick] (\x,\y) circle (.2);
    )
    )
    \foreach \y in {0.5,1.5,2.5}(
    \foreach \x in {0,1,2}(
    \draw[thick,fill = black!40!white] (\x,\y) circle (.2);
    )
    )
 \pgfmathsetmacro{\scaler}{.7}
    \node[scale = \scaler] at (.5,0) {$1$};
    \node[scale = \scaler] at (1.5,0) {$x$};
    \node[scale = \scaler] at (2.5,0) {$x^2$};

    \node[scale = \scaler] at (.5,1) {$y$};
    \node[scale = \scaler] at (1.5,1) {$yx$};
    \node[scale = \scaler] at (2.5,1) {$yx^2$};

    \node[scale = \scaler] at (.5,2) {$y^2$};
    \node[scale = \scaler] at (1.5,2) {$y^2x$};
    \node[scale = \scaler] at (2.5,2) {$y^2x^2$};

    \node[scale = \scaler,color = white] at (0,0.5) {$1$};
    \node[scale = \scaler,color = white] at (1,0.5) {$x$};
    \node[scale = \scaler,color = white] at (2,0.5) {$x^2$};

    \node[scale = \scaler,color = white] at (0,1.5) {$y$};
    \node[scale = \scaler,color = white] at (1,1.5) {$yx$};
    \node[scale = \scaler,color = white] at (2,1.5) {$yx^2$};

    \node[scale = \scaler,color = white] at (0,2.5) {$y^2$};
    \node[scale = \scaler,color = white] at (1,2.5) {$y^2x$};
    \node[scale = \scaler,color = white] at (2,2.5) {$y^2x^2$};

    \node[scale = \scaler,color = zcheckcolor] at (0.5,0.5) {$1$};
    \node[scale = \scaler,color = zcheckcolor] at (1.5,0.5) {$x$};
    \node[scale = \scaler,color = zcheckcolor] at (2.5,0.5) {$x^2$};

    \node[scale = \scaler,color = zcheckcolor] at (0.5,1.5) {$y$};
    \node[scale = \scaler,color = zcheckcolor] at (1.5,1.5) {$yx$};
    \node[scale = \scaler,color = zcheckcolor] at (2.5,1.5) {$yx^2$};

    \node[scale = \scaler,color = zcheckcolor] at (0.5,2.5) {$y^2$};
    \node[scale = \scaler,color = zcheckcolor] at (1.5,2.5) {$y^2x$};
    \node[scale = \scaler,color = zcheckcolor] at (2.5,2.5) {$y^2x^2$};

    \foreach \y in {0,1,2}(
    \foreach \x in {0,1,2}(
    \draw[fill = white,color = white] (\x,\y) circle (.15);
    )
    )

    \node[scale = \scaler,color = xcheckcolor] at (0,0) {$1$};
    \node[scale = \scaler,color = xcheckcolor] at (1,0) {$x$};
    \node[scale = \scaler,color = xcheckcolor] at (2,0) {$x^2$};

    \node[scale = \scaler,color = xcheckcolor] at (0,1) {$y$};
    \node[scale = \scaler,color = xcheckcolor] at (1,1) {$yx$};
    \node[scale = \scaler,color = xcheckcolor] at (2,1) {$yx^2$};

    \node[scale = \scaler,color = xcheckcolor] at (0,2) {$y^2$};
    \node[scale = \scaler,color = xcheckcolor] at (1,2) {$y^2x$};
    \node[scale = \scaler,color = xcheckcolor] at (2.01,2) {$y^2x^2$};
\end{tikzpicture}
\caption{}
\end{subfigure}
\begin{subfigure}[b]{0.45\textwidth}
    \centering
\begin{tikzpicture}[scale = 1.5]
\pgfmathsetmacro{\qubitsize}{.18}
\pgfmathsetmacro{\scaler}{.7}
    \draw[thick] (0,0) -- (2.75,0);
    \draw[thick] (0,1) -- (2.75,1);
    \draw[thick] (0,2) -- (2.75,2);

    \draw[thick] (0,0) -- (0,2.75);
    \draw[thick] (1,0) -- (1,2.75);
    \draw[thick] (2,0) -- (2,2.75);

    \foreach \x in {0.5,1.5,2.5}(
    \foreach \y in {0,1,2}(
    \draw[fill = white,thick] (\x,\y) circle (\qubitsize);
    )
    )
    \foreach \y in {0.5,1.5,2.5}(
    \foreach \x in {0,1,2}(
    \draw[fill = black!40!white,thick] (\x,\y) circle (\qubitsize);
    )
    )
    \draw[fill = xcheckcolor,thick] (2,1.5) circle (\qubitsize);
    \draw[thick, fill = xcheckcolor] (2,.5) circle (\qubitsize);
    \draw[thick, fill = xcheckcolor] (2.5,1) circle (\qubitsize);
    \draw[thick, fill = xcheckcolor] (1.5,1) circle (\qubitsize);
    \draw[thick, fill = zcheckcolor] (1,1.5) circle (\qubitsize);
    \draw[thick, fill = zcheckcolor] (0,1.5) circle (\qubitsize);
    \draw[thick, fill = zcheckcolor] (.5,1) circle (\qubitsize);
    \draw[thick, fill = zcheckcolor] (.5,2) circle (\qubitsize);

    \node[scale = \scaler] at (.5,1) {$y$};
     \node[scale = \scaler] at (1.5,1) {$yx$};
     \node[scale = \scaler] at (2.5,1) {$yx^2$};

     \node[scale = \scaler] at (.5,2) {$y^2$};

    \node[scale = \scaler,color = white] at (2,0.5) {$x^2$};

    \node[scale = \scaler,color = white] at (0,1.5) {$y$};
    \node[scale = \scaler,color = white] at (1,1.5) {$yx$};
    \draw[fill = white,color = white] (2,1) circle (.175);
    \node[scale = \scaler,color = white] at (2,1.5) {$yx^2$};

            \foreach \x in {0.5,1.5,2.5}(
    \foreach \y in {0.5,1.5,2.5}(
    \draw[fill = white,color = white,opacity = .8] (\x,\y) circle (.2);
    )
    )
      \node[scale = \scaler,color = xcheckcolor] at (2,1) {$yx^2$};
          \node[scale = \scaler,color = zcheckcolor] at (0.5,1.5) {$y$};
\end{tikzpicture}
\caption{}
\end{subfigure}
    \caption{The toric code as the two block group algebra code, see \Cref{sec:groupalgebracodedefinition}, associated with $c = 1 + x $ and $d = 1 + y $ in $R = \mathbb{F}_2[x,y]/(x^3 - 1,y^3 - 1)$. It is also an example of a bivariate bicycle code, see \Cref{sec:bbcodedefinition}. In (a), we see that the horizontal qubits (in white) and vertical qubits (in grey) are labeled by monomials (which form the group $\mathbb{Z}/\ell \times \mathbb{Z}/m$). Each monomial also induces an $X$-type stabilizer and a $Z$-type stabilizer as in~\eqref{eq:checksofgroupalgebracodes}. In (b), we depict examples of those, namely $X_{yx^2}$ and $Z_{y}$.}
    \label{fig:toriccode}
\end{figure}
For each $h\in G$ there is an $X$- and $Z$-check defined by
\begin{align}\label{eq:checksofgroupalgebracodes}
X_h=\prod_{\mathclap{g\in \op{supp}(c)}} X_{(hg^{-1})_{\dhor}} \prod_{\mathclap{g\in \op{supp}(d)}} X_{(hg^{-1})_{\dver}} \text{ and }
Z_h=\prod_{\mathclap{g\in \op{supp}(d)}} Z_{(hg)_{\dhor}} \prod_{\mathclap{g\in \op{supp}(c)}} Z_{(hg)_{\dver}}.
\end{align}

\subsection{$Z\!X$-dualities} We recall the framework of \emph{$Z\!X$-dualities} from \cite{breuckmannFoldTransversalCliffordGates2024}, which will allow us to construct fold-transversal gates of the group algebra code $\Code(c,d).$

An \emph{automorphism} of the code $\Code(c,d)$ is a basis preserving automorphism of the underlying chain complex. We denote the group of automorphisms by $\op{Aut}(\Code(c,d)).$
A \emph{$Z\!X$-duality} on $\Code(c,d)$ is a basis preserving isomorphism between the complexes $\Code(c,d)$ and its dual/transposed complex $\Code(c,d)^*.$ We denote the set of $Z\!X$-dualities by $\mathcal{D}_{Z\!X}(\Code(c,d)).$

In fact, the dual code is $\Code(c,d)^*=\Code(\iota(d),\iota(c))$ where 
$$\iota: R\to R, \sum_{g\in G} a_gg\mapsto \sum_{g\in G} a_gg^{-1}.$$
since transpose of multiplication with $r\in R$ is multiplication with $\iota(r).$

We define the \emph{standard $Z\!X$-duality} $\tau_0:\Code(c,d)\to \Code(c,d)^*$ via
\[\begin{tikzcd}[column sep=40pt]
	R & {R_{\dhor}\oplus R_{\dver}} && R \\
	R & {R_{\dhor}\oplus R_{\dver}} && R
	\arrow["\iota"', from=1-1, to=2-1]
	\arrow["{(c,d)}"', from=1-2, to=1-1]
	\arrow["\sigma"', from=1-2, to=2-2]
	\arrow["\begin{array}{c} \left(\begin{smallmatrix} d \\ c \end{smallmatrix}\right) \end{array}"', from=1-4, to=1-2]
	\arrow["\iota", from=1-4, to=2-4]
	\arrow["{(\iota(d),\iota(c))}"', from=2-2, to=2-1]
	\arrow["\begin{array}{c} \left(\begin{smallmatrix} \iota(c) \\ \iota(d) \end{smallmatrix}\right) \end{array}"', from=2-4, to=2-2]
\end{tikzcd}\]
where the map $\sigma$ is defined by
$$\sigma: R_{\dhor}\oplus R_{\dver}\to R_{\dhor}\oplus R_{\dver}, (f,g)\mapsto (\iota(g),\iota(f)).$$
One can compose $Z\!X$-dualities with  automorphisms and by \cite[Lemma 1]{breuckmannFoldTransversalCliffordGates2024}
$$\tau_0\op{Aut}(\Code(c,d))=\mathcal{D}_{Z\!X}(\Code(c,d)).$$
For example, multiplication by $t\in G$ induces an automorphism of the code $\Code(c,d)$ and we obtain $ZX$-duality $\tau_0t$ which acts on physical qubits via $$\tau_0t(g_{\dhor})=((tg)^{-1})_{\dver} \text{ and } \tau_0t(g_{\dhor})=((tg)^{-1})_{\dhor}.$$
\begin{remark}
    \label{rmk:zxtopassfromztox} The $Z\!X$-duality $\tau_0$ exchanges logical $Z$- and $X$-operators. Hence, all properties for logical $Z$-operators translate to $X$-operators. For example, if the code is pure and principal, also the $X$-operators are generated by a single horizontal and a single vertical $X$-operator.
\end{remark}

In \cite{breuckmannFoldTransversalCliffordGates2024} fold-transversal gates are constructed from $Z\!X$-dualites if certain additional assumptions are fulfilled.
Based on their ideas, we now explain how to construct three types of transversal gates that are adapted to group algebra codes.
\subsection{Swap-type gates}
For physical qubits $x,y$ denote by $\op{SWAP}_{x,y}$ the swap gate. Let $\phi$ be any automorphism of the code. Then we obtain a \emph{swap-type gate}
$$\op{SWAP}_{\phi}=\prod_{g\in G}(\op{SWAP}_{g_{\dhor},\phi(g_{\dhor})} \op{SWAP}_{g_{\dver},\phi(g_{\dver})}).$$
We record the following immediate statement.
\begin{theorem}
    The gate $\op{SWAP}_{\phi}$ is an encoded logical gate of the CSS code $\Code(c,d).$
\end{theorem}
\subsection{Hadamard-type gates}
The \emph{Hadamard-type fold-transversal gate} associated to the standard $Z\!X$-duality $\tau_0$ is defined by
$$H_{\tau_0}=\prod_{g\in G}\op{SWAP}_{g_{\dhor},(g^{-1})_{\dver}}\prod_{g\in G} H_{g_\dhor} H_{g_\dver}.$$
\begin{theorem}
    The gate $H_{\tau_0}$ is an encoded logical gate of the CSS-code $\Code(c,d).$
\end{theorem}
\begin{proof} This is \cite[Theorem 6]{breuckmannFoldTransversalCliffordGates2024}.
\end{proof}
We remark, that one may associate a Hadamard-type gate to any $Z\!X$-duality.
\subsection{Phase-type gates}
\begin{figure}
\begin{subfigure}[b]{0.45\textwidth}
    \centering
\begin{tikzpicture}[scale = .8]


\draw[ultra thick, color = green!60!black] (8,-1.5) -- (.5,6);

        \foreach \x in {1,2,3,4,5,6,7} (
            \draw[very thick, color = black!60!white] (\x,-1) -- (\x, 6);
        )
        \foreach \x in {-1,0,1,2,3,4,5} (
            \draw[very thick, color = black!60!white] (1,\x) -- (8,\x);
        )

        \foreach \x in {1,2,3,4,5,6,7} (
            \foreach \y in {-1,0,1,2,3,4,5} (
                \draw[thick, fill = white] (\x + .5, \y) circle (0.2);
            )
        )
        \foreach \x in {1,2,3,4,5,6,7} (
            \foreach \y in {0,1,2,3,4,5,6} (
                \draw[thick, fill = black!20!white] (\x , \y - .5) circle (0.2);
            )
        )


\foreach \x in {0,1,2,3,4,5,6}(
\node[scale = .6, color = green!60!black] at (7 - \x,-.5 + \x) {$S^\dagger$};
)
\foreach \x in {-1,0,1,2,3,4,5}(
\node[scale = .6, color = green!60!black] at (6.5 - \x,  \x) {$S$};
)

\draw[ultra thick, color = green!60!black] (3,4.5) .. controls (2.2,4.3) .. (2,3.5);
\node[scale = .6,color = green!60!black] at (2.5,4.6) {$\op{CZ}$};

\draw[ultra thick, color = green!60!black] (4.5,0) .. controls (5.9,.6) .. (6.5,2);
\node[scale = .6,color = green!60!black] at (6.55,1.5) {$\op{CZ}$};
 \end{tikzpicture} 
 \caption{The gate $\op{CZ}_{\tau_0\omega}$ acts via $\op{CZ}$ on qubits reflected along the diagonal green line as well as $S$ and $S^\dagger$ on the horizontal and vertical qubits on the diagonal.}\label{subfig: czillustration}
 \end{subfigure}
 \hfill
 \begin{subfigure}[b]{0.45\textwidth}
    \centering
 \begin{tikzpicture}[scale = .8]

\pgfmathsetmacro{\zstabxposition}{2}
\pgfmathsetmacro{\zstabyposition}{0}

\pgfmathsetmacro{\xstabxposition}{6}
\pgfmathsetmacro{\xstabyposition}{4}

\draw[ultra thick, color = green!60!black] (8,-1.5) -- (.5,6);
    \draw[fill = zcheckcolor] (\zstabxposition,\zstabyposition) rectangle ++(1,1);

        \foreach \x in {1,2,3,4,5,6,7} (
            \draw[very thick, color = black!60!white] (\x,-1) -- (\x, 6);
        )
        \foreach \x in {-1,0,1,2,3,4,5} (
            \draw[very thick, color = black!60!white] (1,\x) -- (8,\x);
        )

        \foreach \x in {1,2,3,4,5,6,7} (
            \foreach \y in {-1,0,1,2,3,4,5} (
                \draw[thick, fill = white] (\x + .5, \y) circle (0.2);
            )
        )
        \foreach \x in {1,2,3,4,5,6,7} (
            \foreach \y in {0,1,2,3,4,5,6} (
                \draw[thick, fill = black!20!white] (\x , \y - .5) circle (0.2);
            )
        )



    \draw[thick, fill = zcheckcolor] (\zstabxposition+1,\zstabyposition + .5) circle (0.2);
    \draw[thick, fill = zcheckcolor] (\zstabxposition + 0,\zstabyposition + 3.5) circle (0.2);
    \draw[thick, fill = zcheckcolor] (\zstabxposition + 0,\zstabyposition + 4.5) circle (0.2);
    \draw[thick, fill = zcheckcolor] (\zstabxposition + .5,\zstabyposition + 1) circle (0.2);
    \draw[thick, fill = zcheckcolor] (\zstabxposition + 3.5,\zstabyposition ) circle (0.2);
    \draw[thick, fill = zcheckcolor] (\zstabxposition +4.5,\zstabyposition ) circle (0.2);

    \draw[fill = xcheckcolor] (\xstabxposition - .2, \xstabyposition - .2) rectangle ++(.4,.4);

    \draw[thick, fill = xcheckcolor] (\xstabxposition  - .5, \xstabyposition ) circle (0.2);
    \draw[thick, fill = xcheckcolor] (\xstabxposition  + .5, \xstabyposition -3) circle (0.2);
        \fill[color  = xcheckcolor] (\xstabxposition  + .3, \xstabyposition -4) -- (\xstabxposition  + .7, \xstabyposition -4) arc(0:180:0.2) --cycle;
    
\draw[thick] (\xstabxposition  + .5, \xstabyposition -4) circle (0.2);

    \draw[thick, fill = xcheckcolor] (\xstabxposition  , \xstabyposition -.5) circle (0.2);
    \draw[thick, fill = xcheckcolor] (\xstabxposition  -3, \xstabyposition +.5) circle (0.2);
    \fill[color  = xcheckcolor] (\xstabxposition  -4.2, \xstabyposition +.5) -- (\xstabxposition - 3.8, \xstabyposition +.5) arc(0:135:0.2) --cycle;
    \draw[thick] (\xstabxposition  -4, \xstabyposition +.5) circle (0.2);
 \end{tikzpicture} 
 \caption{Conjugation with $\op{CZ}_{\tau_0\omega}$ preserves the stabilizer group. It maps an $X$-stabilizer (in red) $X
_h$ to the product with a $Z$-stabilizer (in blue) $X
_hZ_{\omega(h^{-1}).}$.}\label{subfig: stabilizersundercz}
 \end{subfigure}
    \caption{The phase-type fold-transversal gate $\op{CZ}_{\tau_0\omega}$ on the $[[98,8,12]]$ BB code, see \cref{fig:77code}. The gate is associated to the $Z\!X$-duality $\tau_0\omega$ exchanging $x$ with $y^{-1}$ and $y$ with $x^{-1}$. This corresponds to folding/reflecting the lattice along the diagonal passing through qubits labeled by $x^ny^{-n}.$ In this example, the fold-transversal gates generate the group $C_2\times\op{Sp}_2(\mathbb{F}_{2^3}).$}  \label{fig:cz}
\end{figure}
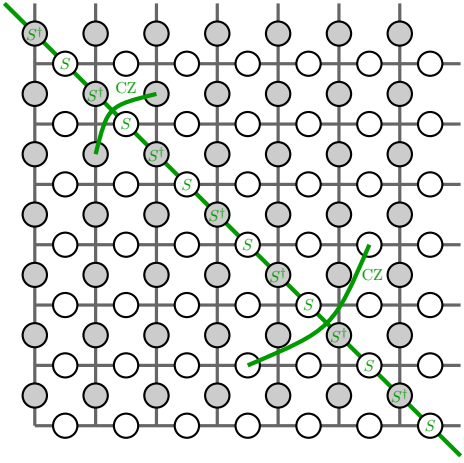
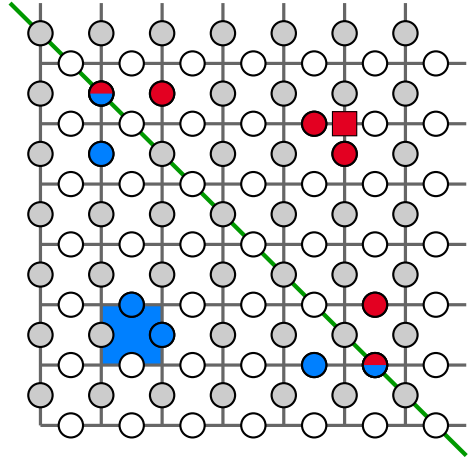

To obtain phase-type gates, we need that the code $\Code(c,d)$ is symmetric in $c$ and $d$ in the following sense.
Assume that there is a group automorphism $\omega$ on $G$ with $\omega^2(g)=g$, such that the induced map on $R$ satisfies $\omega(c)=d.$ 

Then, $\omega$ yields an automorphism of the code, exchanging vertical and horizontal qubits and applying $\omega$ to them. 
We obtain a $Z\!X$-duality $\tau_0\omega$ which acts on physical qubits via $$\tau_0\omega(g_{\dhor})=(\omega(g)^{-1})_{\dhor} \text{ and } \tau_0\omega(g_{\dver})=(\omega(g)^{-1})_{\dver}.$$
Following \cite[Section 2.4.2]{breuckmannFoldTransversalCliffordGates2024} we define the \emph{phase-type fold-transversal gate} associated to the $Z\!X$-duality $\tau_0\omega$ by
$$\op{CZ}_{\tau_0\omega}=\prod_{g\in G_0}S_{g_\dhor}S_{g_\dver}^\dagger\prod_{g\in G_1}\op{CZ}_{g_{\dhor},(\omega(g)^{-1})_{\dhor}} \op{CZ}_{g_{\dver},(\omega(g)^{-1})_{\dver}}.$$
Here, $G_0$ is the set of fixed points and $G_1$ a set of representatives of the two-element orbits under the involution $g\mapsto \omega(g)^{-1}.$
\begin{theorem}
The gate $\op{CZ}_{\tau_0\omega}$ is an encoded logical gate of the CSS code $\Code(c,d).$
\end{theorem}
\begin{proof} We need to show that conjugation with $\op{CZ}_{\tau_0\omega}$ preserves the stabilizer group. Since $\op{CZ}$ commutes with $Z$ is suffices to show that for each $h\in H$ the operator $\op{CZ}_{\tau_0\omega}X_h\op{CZ}_{\tau_0\omega}^\dagger$ is a product of stabilizers. Denote $X_{h,\dhor}$, $X_{h,\dver}$ (and similiarly for $Z$) the horizontal and vertical part of the stabilizers.
We have
\begin{align*}
    \op{CZ}_{\tau_0\omega}X_{h,\dhor}\op{CZ}_{\tau_0\omega}^\dagger
    &=i^{\sigma_\dhor}\prod_{\mathclap{g\in \op{supp}(c)}} X_{(hg^{-1})_{\dhor}}Z_{(\omega(hg^{-1})^{-1})_{\dhor}}
    =(-1)^{\epsilon_\dhor} i^{\sigma_\dhor}X_{h,\dhor}Z_{\omega(h)^{-1},\dhor}
    \text{ and}\\
   \op{CZ}_{\tau_0\omega}X_{h,\dver}\op{CZ}_{\tau_0\omega}^\dagger
   &=(-i)^{\sigma_\dver}\prod_{\mathclap{g\in \op{supp}(d)}} X_{(hg^{-1})_{\dver}}Z_{(\omega(hg^{-1})^{-1})_{\dver}}
    =(-1)^{\epsilon_\dver} (-i)^{\sigma_\dver} X_{h,\dver}Z_{\omega(h)^{-1},\dver}.
\end{align*}
Here, we order the product in the first and second equation via orderings $\op{supp}(c)=\{g_1,\dots,g_w\}$ and $\op{supp}(d)=\{\omega(g_w),\dots,\omega(g_1)\}.$ The signs arise from anti-commuting $Z$'s to the right. One may check that this happens the same number of times on the horizontal and verital part, so $\epsilon_\dhor=\epsilon_\dver$. Similarly, the factors of $i$ and $(-i)$ arise from the applications of $S$- and $S^\dagger$-gates for each fixed point of $\tau_0\omega$ in the support of $X_{h,\dhor}$ and $X_{h,\dver}$, respectively. Again, one may show that $\sigma_\dhor=\sigma_\dver.$

So, we obtain
\begin{align*}
\op{CZ}_{\tau_0\omega}X_{h}\op{CZ}_{\tau_0\omega}^\dagger 
&= (-1)^{\epsilon_\dhor} i^{\sigma_\dver}X_{h,\dhor}Z_{\omega(h)^{-1},\dhor} (-1)^{\epsilon_\dver}(-i)^{\sigma_\dver} X_{h,\dver}Z_{\omega(h)^{-1},\dver} \\
&= X_{h}Z_{\omega(h)^{-1}}
\end{align*}
which shows the statement.
\end{proof}
\section{Logical operators and gates for BB codes}\label{sec:BBcodes}
In this section, we apply the results of \Cref{sec:homologicalalgebraofcodesfromcummutingoperators,sec:groupalgebracodes} in the the special case of \emph{bivariate bicycle (BB) codes} \cite{PhysRevA.88.012311}. We describe the pure logical operators of these codes explicitly. Moreover, we discuss fold-transversal gates.
\subsection{Bivariate Bicycle Quantum Codes} \label{sec:bbcodedefinition}
BB codes are group algebra codes for the product of $G=\mathbb Z/\ell\times \mathbb Z/m$ and were introduced in \cite{PhysRevA.88.012311}. The group algebra of $G$ can be identified a quotient of a bivariate polynomial ring
$$R=\ftwo[G]=\ftwo[x,y]/(x^\ell-1,y^m-1)$$
equipped with the standard basis $\{x^iy^j\}$ where $i=0,\dots ,\ell-1$ and $j=0,\dots,m-1.$

For any choice $c,d\in R$, we obtain an associated BB code  as the group algebra code $\Code(c,d)$, see \Cref{sec:groupalgebracodedefinition}. Similar to the toric code, we may visualize the physical qubits as the vertical and horizontal edges on a $\ell\times m$-grid with periodic boundary. Each vertex and face yields an $X$- and $Z$-stabilizer, respectively, whose support are described by the polynomials $c$ and $d$, see \Cref{fig:77code}.

This way, multiplication with $x$ and $y$ corresponds to shifting an operator on the lattice to the left and bottom, respectively.
\begin{example}\label{ex:somebivariatecodes}The following families of codes are examples of BB codes.
\begin{enumerate}
\item{(Toric code)}
The complex $\Code(1+x,1+y)$ is the chain complex of the toric code on an $\ell\times m$-lattice, see \Cref{fig:toriccode}. The spaces in degree $0,1$ and $2$ can be identified with the free $\ftwo$-vector spaces of vertices, vertical/horizontal edges and faces.
This justifies our vertical/horizontal naming convention.
\item{(Tensor product of cyclic codes)} For univariate polynomials $c=c(x)$ and $d=d(y)$, the complex $\Code(c,d)$ corresponds to the tensor product code (=hypergraph product code) of the two classical cyclic codes with check polynomials $c$ and $d,$ respectively.
\item{(Bicycle codes)} If $m=1$ and hence $R=\ftwo[x,y]/(x^\ell-1)$, for $c=c(x)$ and $d=c(x^{-1})$ one obtains the \emph{bicycle codes} introduced in \cite[Section 4.6]{mackaySparsegraphCodesQuantum2004}.
\item{(Small LDPC codes)} The codes constructed in \cite{bravyi2023highthreshold} show that BB codes can have very favorable properties in small dimensions. We refer to \Cref{tab:properties} where we also list some interesting new parameters found by a computer search.

\item{(Honeycomb color code)} It turns out that the honeycomb color code is exactly the BB code induced by polynomials $c = 1 + x + xy$ and $d = 1 + y + xy$ for $\ell,m$ divisible by three.
\end{enumerate}
\end{example}
\begin{remark} BB codes also admit a description via circulant matrices. Denote by $S_k\in \ftwo^{k\times k}$ the circulant matrix with entries $(S_k)_{i,j}=\delta_{i,j-1}$ where we calculate indices modulo $k.$
The minimal polynomial of $S_k$ is $x^k-1.$
We hence obtain an injective ring homomorphism
$$\phi: R=\ftwo[x,y]/(x^\ell-1,y^m-1)\to \ftwo^{\ell m\times \ell m}, x\mapsto S_\ell\otimes I_m, y\mapsto I_k\otimes S_m$$
which allows us to identify elements in $R$ with sums of tensor products of circulant matrices.
Let $A=\phi(c)$ and $B=\phi(d).$
Under this identification the chain complex defining the code $\Code(c,d)$ becomes
\[\begin{tikzcd}[row sep=small]
	{\ftwo^{\ell m}} & {\ftwo^{\ell m}\oplus \ftwo^{\ell m}} && {\ftwo^{\ell m}}
	\arrow["{(A,B)}"', from=1-2, to=1-1]
	\arrow["\begin{array}{c} \left(\begin{smallmatrix} B \\ A \end{smallmatrix}\right) \end{array}"', from=1-4, to=1-2]
\end{tikzcd}\]
and the associated parity check matrices of the quantum code are
$$H_X=(A,B)\text{ and }H_Z=(B^{\op{tr}}, A^{\op{tr}}).$$
\end{remark}
\subsection{Logical operators}\label{sec:bblogicals}
We now describe the logical operators of BB codes $\Code(c,d)$. This description depends on whether $\ell$ and $m$ are even or odd. We focus mostly on the odd case.
\subsubsection{The odd case}\label{sec:oddlogicals}
The following observation makes the case that  $\ell$ and $m$ are \emph{both odd numbers} particularly easy to handle.
\begin{proposition}\label{prop: oddcase} For $\ell$ and $m$ odd, the ring $R$ is a direct sum of fields. Explicitly, there is an isomorphism
\begin{align}\label{eq:descriptionofR}
    R\cong \bigoplus_{d|\ell,d'|m}(\mathbb{F}_{2^{\op{lcm}(n(d),n(d'))}})^{\gcd(n(d),n(d'))n(d)n(d')}.
\end{align}
\end{proposition}
\begin{proof}
    That $R$ is a direct sum of fields follows from the Artin--Wedderburn theorem since $R$ is semisimple as group algebra of a commutative group $\mathbb{Z}/\ell\times \mathbb{Z}/m$ of odd order. Explicitly, one may show that
$$\ftwo[x]/(x^\ell-1)\cong \bigoplus_{d|\ell} (\mathbb{F}_{2^d})^{n(d)} \text{ and } \ftwo[x]/(x^m-1)\cong \bigoplus_{d|\ell}(\mathbb{F}_{2^d})^{n(d)}.$$
Here $n(d)=\varphi(d)/\op{ord}_2(d)$ where $\varphi(d)$ denotes Euler's totient function and $\op{ord}_2(d)$ is the smallest non-negative integer such that $2^{\op{ord}_2(d)}=1$ modulo $d.$ Now, $R$ is a tensor product of these rings. One may show that for integers $a,b$ there is an isomorphism 
$$\mathbb{F}_{2^a}\otimes \mathbb{F}_{2^b}\cong (\mathbb{F}_{2^{\op{lcm}(a,b)}})^{\gcd(a,b)}$$
which implies the description of $R.$
\end{proof}
\begin{corollary}\label{cor:oddispure}
Let $\ell$ and $m$ be odd. Then, the code $\Code(c,d)$ is pure and principal. The ring $R/(c,d)$ is a product of finite fields of the form $\mathbb{F}_{2^{n_i}}$ for some list of non-negative integers $(n_1,\dots, n_s).$ In particular, one obtains
\begin{align}\label{eq:finitefielddescription}
    H\cong\bigoplus_{i=1}^s \mathbb{F}_{2^{n_i}}\oplus \bigoplus_{i=1}^s \mathbb{F}_{2^{n_i}}
\end{align}
where the two copies correspond to vertical and horizontal logical operators, respectively. 
\end{corollary}
\begin{proof}
 This follows from \Cref{cor:prodoffieldspure,cor:prodoffieldsiso,prop: oddcase}.
\end{proof}

We now describe how to determine the logical operators in practice.
For this, consider the so-called \emph{$2D$-cyclic codes}
$$\ann{c}\subset R \text{ and } \ann{d}\subset R$$
whose code words consist of all $r\in R$ for which $cr=0$ and $dr=0,$ respectively. Then, one may always find generator polynomials $G\in R$ and $H\in R$ such that all code words of the $2D$-cyclic codes can be obtained by sums of shifts of $G$ and $H$ on the lattice, that is,
$$\ann{c}=(P) \text{ and } \ann{d}=(Q)$$
where for $f\in R$ we denote by $(f)=\{\sum a_{i,j}x^iy^i f\}$ the ideal generated by $f.$ While finding such $P$ and $Q$ is always possible theoretically, 
in \Cref{sec:semiperiodic} we consider a special family of codes, for which there is an explicit algorithm.
We now obtain two logical $Z$-operators of the BB code: the first operator is supported on the vertical qubits according to the support of $P$, the second operator is supported on the horizontal qubits according to the support of $Q.$ All other logical $Z$-operators can be obtained by products of shifts of these operators.
We obtain that the following map is an isomorphism
\begin{equation}\label{eq: isomorphismoddcase}
  R/(c,d)\oplus R/(c,d)\to H, ([f],[g])\mapsto [fP,gQ].  
\end{equation}

\subsubsection{The even case} The situation is more difficult $\ell$ or $m$ is even. In this case, the ring $R$ is non-reduced and the code $\Code(c,d)$ is not pure or principal in general: both properties depend on the choice of polynomials $c$ and $d.$ We refer to Table~\ref{tab:properties} for some (non)-examples.

However, there are some special cases where one can say more. The first case are HGPs of cyclic codes, see also~\cite{Quintavalle_2023}.
\begin{proposition}
     If both $c=c(x)$ and $d=d(y)$ are univariate polynomials in $x$ and $y$ each, then $\Code(c,d)$ is pure.
\end{proposition}
Another interesting case is when the code is pure and the polynomials $c$ and $d$ are of a special type.
\begin{definition}\label{def:semiperiodicpolynomial}
We call $c\in R$
semiperiodic if $c(x,y)=x^k+\zeta(y)$ for some polynomial $\zeta(y)$ in $y$ and $kk'=\ell.$
\end{definition}
It turns out that bivariate bicycle codes defined by semi-periodic polynomials have especially nice logical operators, see~\Cref{sec:semiperiodic} for a discussion.

\begin{theorem}
    If the bivariate bicycle code is pure, then it is principal if both polynomials $c,d$ are each either univariate or semiperiodic.
\end{theorem}
\begin{proof}
In the first case, if $c(x,y)=c(x)$ let $g(x)$ be a generator polynomials of the cyclic code with check polynomial $c(x)$. Then $\ann{c}=(g)\subset R$ is principal. The same works with $d$ instead of $c$.
In the second case, this follows from \Cref{thm:semiperiodic}.
\end{proof}
\begin{remark}
    Write $k=k'k''$ for $k'$ odd and $k''$ a power of two and let $\underline{x}=x-1$. Then the Chinese remainder theorem shows that
    $$\ftwo[\underline{x}]/(x^{k}-1)=\ftwo[x]/(x^{k'}-1)\oplus \ftwo[\underline{x}]/(\underline{x}^{k''}).$$
    The same idea applies $y$ and $\ell$. So, the even case reduces to the odd case and the `nilpotent' case $\ftwo[\underline{x}]/(\underline{x}^{k''})$. We leave the nilpotent case to the interested reader.
\end{remark}
\subsection{Fold-transversal gates} \label{sec:bbgates}
Since a BB code is an example of a group algebra code, we can apply our results on fold-transversal gates in \Cref{sec:groupalgebracodes} to construct fold-transversal gates with no overhead. Let $k=\dim H$ denote the number of logical qubits. Then transversal gates are elements of the Clifford group $\mathcal{C}_k.$ It is sometimes convenient to calculate modulo elements of the Pauli group and we study a gate, say $U$, in terms of its co-set $[U]$ in the symplectic group
$$\mathcal{C}_k/\mathcal{P}_k=\op{Sp}(H\oplus H^*)=\op{Sp}_{2k}(\ftwo).$$

First, we always obtain the swap-gates 
$$\op{SWAP}_x,\op{SWAP}_y$$
that arise by shifting the lattice in the $x$ and $y$ direction, respectively.
In the case that $m,\ell$ are odd, we explained in \Cref{sec:oddlogicals} that the space of logical operators $H$ can be decomposed into a direct sum of fields of the form $\mathbb{F}_{2^{n_i}}$, see \eqref{eq:finitefielddescription}. With respect to this description, the gates act via multiplication with an $\ell$-th and $m$-th root of unity, respectively, on the individual fields. Moreover, these gates preserve vertical and horizontal logical operators. Hence, they yield elements of qudit-Clifford gates and their representatives are in a subgroup
$$\langle [\op{SWAP}_x],[\op{SWAP}_y] \rangle \subset \prod_{i=1}^s\op{Sp}_2(\mathbb{F}_{2^{n_i}})\subset \op{Sp}_{k}(\mathbb{F}_2) \times \op{Sp}_{k}(\mathbb{F}_2) \subset \op{Sp}_{2k}(\mathbb{F}_2).$$
Another fold-transversal gate arises from the $Z\!X$-duality $\tau_0$ which exchanges $x$ and $x^{-1}$, $y$ and $y^{-1}$. It induces the Hadamard-type gate $H_{\tau_0}.$ This gate exchanges vertical and horizontal as well as logical $Z$- and $X$-operators.

To obtain more gates, we will focus on codes that fulfill the following property.
\begin{definition}\label{def:symmetric}
    A bivariate bicycle code $\Code(c,d)$ is called \emph{symmetric} if $\ell=m$ and $c(x,y)=d(y,x).$
\end{definition}
In this case, we obtain an automorphism $\omega$ of the code which exchanges vertical and horizontal directions as well as $x$ and $y.$ This yields another swap-type gate $\op{SWAP}_\omega$. 

Moreover, we consider the $Z\!X$-duality $\tau_0\omega$ which exchanges $x$ by $y^{-1}$ and $y$ by $x^{-1}$. It induces a phase-type gate $\op{CZ}_{\tau_0\omega}$ which acts as a $\op{CZ}$-gate preserving vertical and horizontal logical operators. The representative hence is an element of the subgroup
$$ [\op{CZ}_{\tau_0\omega}] \in \op{Sp}_{k}(\mathbb{F}_2) \times \op{Sp}_{k}(\mathbb{F}_2) \subset \op{Sp}_{2k}(\mathbb{F}_2).$$
So, in total, we obtain for symmetric $BB$-codes the group of fold-transversal gates
$$G=\langle[\op{Swap}_x], [\op{Swap}_y], [\op{Swap}_\omega],[H_{\tau_0}],[\op{CZ}_{\tau_0\omega}]  \rangle\subset \op{Sp}_{2k}(\ftwo).$$

\subsection{Semiperiodic two-dimensional cyclic codes}\label{sec:semiperiodic}
In the following, we want to determine an explicit construction for code words of certain classical binary two-dimensional cyclic codes. This will be useful for the explicit examples we will present later.
Again, let $R=\ftwo[x,y]/(x^\ell-1,y^m-1).$ Ideals in $R$ are called two-dimensional cyclic codes, see \cite{IMAI19771}. In this section, we want to explicitly describe the code words for a special family of such codes.
\begin{definition}\label{def:semiperiodic}
We call
$\ann{c(x,y)}\subset R$ a \emph{semiperiodic two-dimensional cyclic code} if $c$ is semiperiodic, see \Cref{def:semiperiodicpolynomial}. That is, $c(x,y)=x^k+\zeta(y)$ for some polynomial $\zeta(y)$ in $y$ and $kk'=\ell.$
\end{definition}
The name stems from the fact that codewords $f(x,y)\in \ann{c(x,y)}$ are semiperiodic by translations by $k$ steps in the $x$-direction
\begin{equation}
    x^kf(x,y)=\zeta(y)f(x,y) \label{eq:semiperiodic}
\end{equation}
We now construct a generator polynomial for a semiperiodic code with check polynomial $c(x,y).$

Let $\chi(y)=\zeta(y)^{k'}-1$, $\underline{\chi}(y)=\gcd(\chi(y),y^m-1)$ and $g(y)\underline{\chi}(y)=y^m-1.$
Hence, $g(y)$ is the generator polynomial of the cyclic code on $\ftwo[y]/(y^m-1)$ defined by the check polynomial $\chi(y)=\zeta(y)^{k'}-1.$ We denote the distance of this cyclic code by $d_{\chi}.$
Lastly, let
$$P(x,y)=\sum_{i=0}^{k'-1}x^{\ell-ik}\zeta(y)^ig(y).$$
\begin{theorem}\label{thm:semiperiodic} The semiperiodic two-dimensional cyclic code has generator polynomial $P(x,y),$ so
$$\ann{c(x,y)}=(P(x,y))\subset R.$$
Moreover, we have obtain $k'm\geq d_c\geq k'd_{\chi}=\frac{\ell}{k}d_{\chi}.$
\end{theorem}
\begin{proof}
Write $ak+b\ell=\underline{k}$ and let $f(x,y)\in \ann{c}.$
Then $$x^kf(x,y)=\zeta(y)f(x,y).$$
By applying the equation $k'$ times and using $x^{k'k}=x^{\ell}=1$ we obtain
$$f(x,y)=\zeta(y)^{k'}f(x,y)\Leftrightarrow f(x,y)(\zeta(y)^{k'}-1)=f(x,y)\chi(y)=0.$$
Now, we write
$f(x,y)=\sum_{j=0}^{\ell-1}x^jf_j(y).$
We obtain, counting indices modulo $\ell$,
$$f_j(y)\chi(y)=0 \text{ and } f_{j-k}(y)=f_j(y)\zeta(y).$$
If follows that there are polynomials $a_j(y)$ such that
$$f_j(y)=a_j(y)g(y)\text{ and }f(x,y)=\sum_{j=0}^{k-1}x^ja_j(y)P(x,y)\in (P(x,y)).$$
We assume that $f(x,y)\neq 0$ and without loss of generality $f_0(y)\neq 0.$ Then also $f_0(y)$, $f_k(y)$, $f_{2k}(y)$,$\dots$, $f_{(k'-1)k}(y)$ are non-zero elements of $\ann{\chi}$ and hence 
\[|f(x,y)|\geq k'd_\chi.\]
This gives the lower bound on $d_c.$ The upper bound comes from $k'm\geq |P(x,y)|.$
\end{proof}
\begin{remark}
    The proof shows that the distance $d_c$ is precisely
    $$d_c=\min\left\{\sum_{i=0}^{k'-1}|\zeta(y)^ia(y)|\,\middle|\, 0\neq a(y)\in \ann{\chi(y)}\right\}$$
    and hence is a quantity that depends purely on properties of the cyclic code $$\ann{\chi(y)}=(g(y))\subset \ftwo[y]/(y^m-1).$$
\end{remark}
\section{Two extended examples} \label{sec:examples}
We discuss two bivariate bicycle codes with favorable properties. We describe how to construct logical operators and fold-transversal gates in detail. Some of the groups were analyzed using GAP~\cite{GAP4}.
\subsection{A $[[98,6,12]]$-code}
In this section we take a closer look at the $[[98,6,12]]$ bivariate bicycle code $\Code(c,d)$ for polynomials $c=x+y^3+y^4$ and $d=y+x^3+x^4$ on a grid of size $\ell\times m=7\times 7$, see \Cref{fig:77code}.

\subsubsection{Constructing logical operators}\label{sec:constructinglogicals77}
We first note that the code is pure and principal by \Cref{cor:oddispure} since $\ell$ and $m$ are odd, that is $H=H_{\dhor}\oplus H_{\dver}.$ So, similarly to the toric code, each logical operator decomposes uniquely as a sum of vertical and horizontal logical operators. Recall from \Cref{eq: isomorphismoddcase} that
\begin{equation*}
  R/(c,d)\oplus R/(c,d)\to H, ([f],[g])\mapsto [fP,gQ]  
\end{equation*}
is an isomorphism, where $P,Q\in R$ are generators of the ideals $\ann{c}$ and $\ann{d}$, respectively.

We now explain how to determine $P$ and $Q$ explicitly. Note that $\ann{c}\subset R$ is a semi-periodic two-dimensional cyclic code, see \Cref{def:semiperiodic}. To determine the code words, we write $c=x+\zeta(y)$, where $\zeta(y)=y^3+y^4.$ We set $$\chi(y)=\zeta(y)^7+1=1+y+y^2+y^3+y^4+y^6+y^7.$$ Then, the classical cyclic code $\ann{\chi(y)}\subset\ftwo[y]/(y^7-1)$ has generator polynomial 
$$g(y)=1+y.$$
By \Cref{thm:semiperiodic}, we can obtain that
$\ann{c}=(P(x,y))$
for
$$
P(x,y)=\sum_{i=0}^6(x^{7-i}\zeta(y)^ig(y)).
$$
See \Cref{subfig: logical operator} for the support of this logical operator.

One may show that $(c,d)=(x-y, x^3+x^2+1),$ so that
$$R/(c,d)\cong \ftwo[x]/(x^3+x^2+1)\cong \mathbb{F}_{2^3}.$$ Hence, we obtain an isomorphism $H_{\dhor}\cong \mathbb{F}_{2^3}$. 

Similarly, $\ann{d}=(Q(x,y))$ where $Q(x,y)=P(y,x)$ and $H_{\dver}\cong \mathbb{F}_{2^3}.$ Hence the space of logical operators can be identified with 
$$H\cong \mathbb{F}_{2^3}\oplus \mathbb{F}_{2^3}.$$
Dual bases of $H$ and $H^*$ are given by
\begin{equation*}
\begin{aligned}
    L_{1,\dhor} &= [P(x,y), 0], & \; L_{2,\dhor} &= [xP(x,y), 0] & \; L_{3,\dhor} &= [x^{3}P(x,y), 0],\\
    L_{1,\dver} &= [0, Q(x,y)], & \; L_{2,\dver} &= [0, xQ(x,y)] & \; L_{3,\dver} &= [0, x^{3}Q(x,y)]\text{ and}\\
    L_{1,\dhor}^* &= [xQ(x^{-1},y^{-1}), 0], & \; L_{2,\dhor}^* &= [Q(x^{-1},y^{-1}), 0] & \; L_{3,\dhor}^* &= [x^{-2}Q(x^{-1},y^{-1}), 0],\\
    L_{1,\dver}^* &= [0, xP(x^{-1},y^{-1})], & \; L_{2,\dver}^* &= [0, P(x^{-1},y^{-1})] & \; L_{3,\dver}^* &= [0, x^{-2}P(x^{-1},y^{-1})].
\end{aligned}
\end{equation*}
With respect to these, multiplication by $x$ (and also $y$) on $H_\dhor$ and $H_\dver$ is given by the matrix 
\begin{equation}\label{eq:matrixX}
T = \begin{pmatrix}
0 & 1 & 0 \\
1 & 0 & 1 \\
0 & 1 & 1
\end{pmatrix}.
\end{equation}
\subsubsection{Fold-transversal gates}\label{sec:gates77} We now describe the action of the fold-transversal gates, as defined in \Cref{sec:groupalgebracodes}, on the logical operators.

The gates are elements of Clifford group $\mathcal{C}_6$. To describe the action in a compact form, we describe them modulo elements of the Pauli group $\mathcal{P}_6.$ 
The quotient $$\mathcal{C}_6/\mathcal{P}_6=\op{Sp}(H\oplus H^*)=\op{Sp}_{12}(\ftwo)$$ is a symplectic group. We hence describe the gates in terms of symplectic matrices with respect to the basis of $H\oplus H^*=H_{\dhor}\oplus H_{\dver}\oplus H_{\dhor}^*\oplus H_{\dver}^*$ defined above.

Multiplication with $x$ yields a swap-type gate with matrix
\[[\op{Swap}_x]=
\left[
\begin{array}{c|c|c|c}
    T & 0 & 0 & 0 \\ \hline
    0 & T & 0 & 0 \\ \hline
    0 & 0 & T^{-\op{tr}} & 0 \\ \hline
    0 & 0 & 0 & T^{-\op{tr}}\\ 
\end{array}
\right]\in \op{Sp}_{12}(\ftwo)
\]
with $X$ as in \eqref{eq:matrixX}. The automorphism $\omega$ yields the swap-type gate
\[[\op{Swap}_\omega]=
\left[
\begin{array}{c|c|c|c}
    0 & I_3 & 0 & 0 \\ \hline
    I_3 & 0 & 0 & 0 \\ \hline
    0 & 0 & 0 & I_3 \\ \hline
    0 & 0 & I_3 & 0\\ 
\end{array}
\right]\in \op{Sp}_{12}(\ftwo)
\]
where $I_3$ denotes the $3\times 3$ identity matrix.

Next, we consider the $Z\!X$-duality $\tau_0$ which exchanges $x$ and $x^{-1}$, $y$ and $y^{-1}$ as well as vertical and horizontal qubits. It induces maps $H_{\dhor}\to H^*_{\dver}$ and $H_{\dver}\to H^*_{\dhor}$ which are both given by the matrix $T.$
The corresponding Hadamard-type gate has matrix
\[[H_{\tau_0}]=
\left[
\begin{array}{c|c|c|c}
    0 & 0 & 0 & T \\ \hline
    0 & 0 & T & 0 \\ \hline
    0 & T^{-\op{tr}} & 0 & 0 \\ \hline
    T^{-\op{tr}} & 0 & 0 & 0\\ 
\end{array}
\right]\in \op{Sp}_{12}(\ftwo).
\]
Lastly, we consider the $Z\!X$-duality $\tau_0\omega$ which exchanges $x$ by $y^{-1}$ and $y$ by $x^{-1}$. It induces maps $H_{\dhor}\to H^*_{\dhor}$ and $H_{\dver}\to H^*_{\dver}$ which are also both given by $T$.
The corresponding phase-type gate has matrix
\[[\op{CZ}_{\tau_0\omega}]=
\left[
\begin{array}{c|c|c|c}
    I_3 & 0 & X & 0 \\ \hline
    0 & I_3 & 0 & X \\ \hline
    0 & 0 & I_3 & 0 \\ \hline
    0 & 0 & 0 & I_3\\ 
\end{array}
\right]\in \op{Sp}_{12}(\ftwo).
\]
In fact, the group generated by these three gates $$G=\langle[\op{Swap}_x], [\op{Swap}_\omega],[H_{\tau_0}],[\op{CZ}_{\tau_0\omega}]  \rangle\subset \op{Sp}_{12}(\ftwo)$$
is isomorphic to $C_2\times\op{PSL}_{2}(\mathbb{F}_{2^3})=C_2\times \op{Sp}_{2}(\mathbb{F}_{2^3})$. Interestingly, $\op{Sp}_{2}(\mathbb{F}_{2^3})$ is the quotient of the generalized Clifford group and Pauli group of a single $\mathbb{F}_{2^3}$-qudit. 

\subsection{A $[[162,8,12]]$-code} We consider the $[[162,8,12]]$ bivariate bicycle code $\Code(c,d)$ for $c=x^3+y+y^2$ and $d=y^3+x+x^2$ on an $\ell\times m=9\times 9.$
\subsubsection{Constructing logical operators}
As in \Cref{sec:constructinglogicals77} $H=H_{\dhor}\oplus H_{\dver}$ and $\ann{c}\subset R$ is a semiperiodic two-dimensional cyclic code where $c=x^3+\zeta(y)$.
We can obtain that
$\ann{c}=(P(x,y))$
for $$P(x,y)=(1+x^3+x^6)(y+y^2)(1+y^3+y^6).$$
Since $(c,d)=(x^2 + x + 1, y^2 + y + 1)$ there is an isomorphism
$$(\mathbb{F}_{2^2})^2\cong \mathbb{F}_{2^2}\otimes \mathbb{F}_{2^2}=R/(c,d)\to H_{\dhor}, [r]\mapsto [rP,0].$$
We obtain a basis of $H_v$ by
\begin{equation*}
\begin{aligned}
    L_{1,\dhor} &= [P(x,y), 0], & \; L_{2,\dhor} &= [xP(x,y), 0], \\
    L_{3,\dhor} &= [yP(x,y), 0], & \quad L_{4,\dhor} &= [xyP(x,y), 0].
\end{aligned}
\end{equation*}
By applying $\omega$, which exchanges $x,y$ and horizontal/vertical, we also obtain a basis of $H_\dver$ and 
$$H\cong (\mathbb{F}_{2^2}\otimes \mathbb{F}_{2^2})\oplus (\mathbb{F}_{2^2}\otimes \mathbb{F}_{2^2}).$$
In this case, a dual basis can be constructed by simply applying to each basis vector $\tau_0\omega$ which exchanges $x$ by $y^{-1}$ and $y$ by $x^{-1}$.

\subsubsection{Fold-transversal gates} 
As in \Cref{sec:gates77} one can compute the subgroup generated by fold-transversal gates
 $$G=\langle[\op{Swap}_x], [\op{Swap}_y], [\op{Swap}_\omega],[H_{\tau_0}],[\op{CZ}_{\tau_0\omega}]  \rangle\subset \op{Sp}_{16}(\ftwo).$$
In this case, there is an isomorphism
$G\cong \op{Sp}_2(\mathbb{F}_{2^2})\times (\op{Sp}_2(\mathbb{F}_{2^2})\rtimes C_2).$ 

In the semi-direct product, the cyclic group in two elements $C_2$ acts on $\op{Sp}_2(\mathbb{F}_{2^2})$ by the Frobenius endomorphism, which squares each entry in the matrix.

One of the copies of $\op{Sp}_2(\mathbb{F}_{2^2})$ in $G$ arises as the subgroup
$$\langle[\op{Swap}_x], [\op{Swap}_\omega][H_{\tau_0}],[\op{CZ}_{\tau_0\omega}]  \rangle\subset G.$$

\section{Conclusion}
In this paper, we have introduced methods from homological and commutative algebra to investigate qLDPC codes. For various families like quantum two-block group algebra codes and bivariate bicycle codes, we used these methods to reveal new structural insights into the logical operators and fold-transversal gates on these codes: In particular, we generalized the Künneth formula for these codes which allowed us to construct logical operators in a structured manner. As a corollary, we demonstrated that BB codes under certain easy-to-check conditions share many of the favorable properties of toric and hypergraph product codes. This includes the ability to generate all logical operators from a single vertical and a single horizontal logical operator. These logical operators correspond to code words of classical two-dimensional cyclic codes, and we provided explicit formulas for them in the semiperiodic case. 

We believe that this new approach to handling qLDPC codes opens the door to a variety of new directions. 
\begin{enumerate}
    \item We have demonstrated how our methods give deep structural insights for quantum two-block group algebra codes and bivariate bicycle codes. We believe that they might also be applicable to other qLDPC code families. For example, for the higher-dimensional generalizations introduced in \Cref{sec:nblockcodes}, we expect that we can implement fold-transversal gates in higher levels of the Clifford hierarchy using our methods.
    \item In this work we saw how one can use the structure of the spaces of logical operators to get a better understanding of which fold-transversal gates can be implemented. We expect that our detailed understanding of the logical operators also helps to perform lattice surgery or execute other proposals of fault-tolerant implementation of gates.
    \item For specific examples of codes, we demonstrated a large family of Clifford gates that can be implemented using the fold-transversal gates explained in this work. It remains to understand if there are even more fold-transversal gates coming from other $ZX$-dualities.
    \item More generally, we gave a concise characterization of when a bivariate bicycle code is pure, that is, has a basis of logical operators supported on horizontal or vertical edges only. We believe that pureness can be useful for various other applications such as decoding.
\end{enumerate}
\bibliographystyle{alpha}
\bibliography{bivariatebicycle}
\clearpage
\appendix
\section{Balanced products and the Künneth spectral sequence}\label{sec:balancedproductkuenneth}
In this section we show that the quantum code $\Code(c,d)$ arises as a \emph{balanced product code}, see \cite{breuckmannBalancedProductQuantum2021b,breuckmannQuantumLowDensityParityCheck2021a}. We show how this provides a conceptual explanation of the fundamental exact sequence \Cref{eq:fundamentalsequence} in terms of the Künneth spectral sequence.

Consider the following complexes (concentrated in degrees $0,1$) induced by multiplication with $c$ and $d,$ respectively,
\[\begin{tikzcd}[row sep=tiny]
	{C_\bullet:R} & R \\
	{D_\bullet:R} & R
	\arrow["c"', from=1-2, to=1-1]
	\arrow["d"', from=2-2, to=2-1]
\end{tikzcd}\]
To this, we associate the tensor product double complex $C_\bullet\boxtimes_R D_\bullet$ which has the form
\[\begin{tikzcd}
	{R\otimes_RR} & {R\otimes_RR} \\
	{R\otimes_RR} & {R\otimes_RR}.
	\arrow["{\id\otimes d}"', from=1-1, to=2-1]
	\arrow["c\otimes\id"', from=1-2, to=1-1]
	\arrow["{\id\otimes d}", from=1-2, to=2-2]
	\arrow["c\otimes\id"', from=2-2, to=2-1]
\end{tikzcd}\]
Passing to the total complex this yields the tensor product chain complex $C_\bullet\otimes_RD_\bullet$ given by
\[\begin{tikzcd}[column sep=large]
	{R\otimes_RR} & {R\otimes_RR\oplus R\otimes_RR} & {R\otimes_RR}
	\arrow["{(c\otimes\id,\id\otimes d)}"', from=1-2, to=1-1]
	\arrow["\begin{array}{c} \left(\begin{smallmatrix}\id\otimes d \\ c\otimes\id \end{smallmatrix}\right) \end{array}"', from=1-3, to=1-2]
\end{tikzcd}\]
If $C_\bullet$ and $D_\bullet$ are chain complexes representing classical linear codes, the quantum code associated to $C_\bullet\otimes_RD_\bullet$ is a \emph{balanced product quantum code}. 

Multiplication $r\otimes s\mapsto rs$ allows to identify $R\otimes_R R$ and $R$ and we obtain that the chain complex/quantum code $\Code(c,d)$, see \Cref{eq:complexofCcd}, arises as tensor/balanced product
$$\Code(c,d)=C_\bullet\otimes_RD_\bullet.$$
This interpretation can be helpful, since the homology of a tensor product is subject to the Künneth spectral sequence (a generalization of the Künneth formula). We explain now how this spectral sequence describes how the homology groups of $C_\bullet\otimes_RD_\bullet$
are `built' from the homology groups of the factors $C_\bullet$ and $D_\bullet.$

The homology of $C_\bullet$ and $D_\bullet$ is concentrated in degrees 0,1 and given by
\begin{align*}
H_0(C_\bullet)=R/(c), H_0(D_\bullet)=R/(d),
H_1(C_\bullet)=\ann{c}&\text{ and }H_1(D_\bullet)=\ann{d}.
\end{align*}
The Künneth spectral sequence, see \cite[Theorem 56.4]{weibelIntroductionHomologicalAlgebra1994}  is of the form  $$E^2_{p,q}=\bigoplus_{k_1+k_2=q}\Tor^R_p(H_{k_1}(C_\bullet), H_{k_2}(D_\bullet))\Rightarrow H_{p+q}(C_\bullet\otimes_R D_\bullet).$$
Now $E^2_{p,q}$ is zero outside of $q=0,1,2$ and $p\geq 0.$ This implies that the spectral sequence degenerates on page $E^3$ and there is a short exact sequence
\begin{equation}\label{sesforH}\begin{tikzcd}[column sep=scriptsize]
	0 & {E_{0,1}^3} & {H} & {E_{1,0}^3} & 0.
	\arrow[from=1-1, to=1-2]
	\arrow[from=1-2, to=1-3]
	\arrow[from=1-3, to=1-4]
	\arrow[from=1-4, to=1-5]
\end{tikzcd}\end{equation}
The entries $E_{p,q}^3$ on the third page $E^3$ arise from the homology of a differential $d^2$ of bidegree $(-2,1)$ between the entries on the second page $E^2.$ Since $E^2_{-1,1}=E^2_{3,-1}=0,$ we obtain
$$E_{1,0}^3=E_{1,0}^2=\Tor^R_1(H_0(C_\bullet),H_0(D_\bullet))=\Tor^R_1(R/(c),R/(d))\cong \frac{(c)\cap(d)}{(cd)}.$$
For the last isomorphism, see \cite[Exercise 3.1.2]{weibelIntroductionHomologicalAlgebra1994}.
Since $E^2_{-2,0}=E^2_{4,-1}=0$, we obtain a short exact sequence
\begin{equation}\label{sesforE}
    \begin{tikzcd}
	0 & {E^2_{2,0}} & {E^2_{0,1}} & {E^3_{0,1}} & 0
	\arrow[from=1-1, to=1-2]
	\arrow["{d^2_{2,0}}", from=1-2, to=1-3]
	\arrow[from=1-3, to=1-4]
	\arrow[from=1-4, to=1-5]
\end{tikzcd}\end{equation}
whose terms we may compute explicitly as
\begin{align*}
E^2_{0,1}&=H_1(C_\bullet) \otimes_R H_0(C_\bullet) \oplus  H_0(C_\bullet) \otimes_R H_1(C_\bullet)\\&=\ann{c}\otimes_RR/(d)\oplus R/(c)\otimes_R\ann{d}\\
&\cong\frac{\ann{c}}{\ann{c}(d)}\oplus\frac{\ann{d}}{(c)\ann{d}} \text{ and}\\
E^2_{2,0}&=\Tor^R_2(H_0(C_\bullet),H_0(D_\bullet))=\Tor^R_2(R/(c),R/(d))\cong \frac{\ann{cd}}{M}.
\end{align*}
where $M$ is defined as in \Cref{thm:fundamentalsequence} and we use \Cref{prop:descriptionoftor2} and \cite[Exercise 3.1.1]{weibelIntroductionHomologicalAlgebra1994}. for the explicit description of the $\Tor_2$-term.
Putting together the short exact sequences in \eqref{sesforH} and \eqref{sesforE} as well as the explicit description for its terms yields the fundamental exact sequence \eqref{eq:fundamentalsequence} from \Cref{thm:fundamentalsequence}.
\section{$n$-fold balanced product quantum codes}\label{sec:nblockcodes}
The balanced product description of the codes $\Code(c,d)$ in \Cref{sec:balancedproductkuenneth} provides an immediate generalization to $n$-fold balanced product quantum codes. Namely, let $c_1,\dots, c_n\in R$. Denote by $C_\bullet^i$ the complex
\[\begin{tikzcd}[row sep=tiny]
	{C_\bullet^{(i)}:R} & R.
	\arrow["c_i"', from=1-2, to=1-1]
\end{tikzcd}\]
Then we obtain a chain complex with $n$ terms in degree $0,\dots,n$ via the iterated balanced product
$$C_\bullet^{(1)}\otimes_R\dots \otimes_R C_\bullet^{(n)}.$$
Hence, we obtain a non-trivial quantum code, which we denote by $\Code(i,c_1,\dots,c_2)$ from the differential starting and ending in the $i$-th position, for $i=1,\dots,n-1.$ We note that the higher and lower differentials yield relations of the $X$- and $Z$-checks, which can be a favorable property for decoding.

We explicitly discuss the example of $n=3$. The associated triple tensor product complex $C_\bullet^{(1)}\boxtimes_RC_\bullet^{(2)} \boxtimes_R C_\bullet^{(3)}$ is a cube
\[\begin{tikzcd}
	&& R & R \\
	R & R & R & R \\
	R & R
	\arrow[from=1-3, to=2-1]
	\arrow["{d}"', from=1-3, to=2-3]
	\arrow["{c}"', from=1-4, to=1-3]
	\arrow[shift left, from=1-4, to=2-2]
	\arrow["{d}", from=1-4, to=2-4]
	\arrow["{d}"', from=2-1, to=3-1]
	\arrow["{c}", from=2-2, to=2-1]
	\arrow["{d}", from=2-2, to=3-2]
	\arrow[shift right, from=2-3, to=3-1]
	\arrow["{c}"', from=2-4, to=2-3]
	\arrow[from=2-4, to=3-2]
	\arrow["{c}", from=3-2, to=3-1]
\end{tikzcd}\]
where we abbreviate $c=c_1$, $d=c_2$ and the arrows without label are multiplication by $e=c_3.$ The resulting total complex is
\newcommand{\matrixA}{\left(\begin{smallmatrix}d & c & e\end{smallmatrix}\right)}
\newcommand{\matrixB}{\left(\begin{smallmatrix}e & 0 & c \\ 0 & e & d \\ d & c & 0\end{smallmatrix}\right)}
\newcommand{\matrixC}{\left(\begin{smallmatrix}c \\ d \\ e\end{smallmatrix}\right)}

\[
\begin{tikzcd}
    R & {R^3} & {R^3} & R
    \arrow["{\matrixA}"', from=1-2, to=1-1]
    \arrow["{\matrixB}"', from=1-3, to=1-2]
    \arrow["{\matrixC}"', from=1-4, to=1-3]
\end{tikzcd}
\]

From the differentials we hence obtain two quantum codes. 
We note that the $n=4$ case is more symmetric, but we leave this to the interested reader.
\section{Quantum codes from bimodules}\label{app:bimodules} We briefly explain how the results from \Cref{sec:quantumcodesfromcommutativerings} generalize to non-commutative rings and modules. Let $R$ be any ring (with unit), and $M$ an $R$-bimodule. This means that $M$ is equipped with a left and right action of $R,$ respectively, and these actions commute, so 
$$r(ms)=(rm)s \text{ for all } r,s\in R \text{ and } m\in M.$$
Then, for fixed $c,d \in R,$ we obtain the following commutative diagram
\begin{equation}\label{eq:doublecomplexofbimodulecodes}
\begin{tikzcd}
	M_{\hor} & M \\
	M & M_{\ver}.
	\arrow["\rmp{d}"', from=1-1, to=2-1]
	\arrow["\lmp{c}"', from=1-2, to=1-1]
	\arrow["\rmp{d}", from=1-2, to=2-2]
	\arrow["\lmp{c}"', from=2-2, to=2-1]
\end{tikzcd}
\end{equation}
where we denote the operators given by left and right multiplication by
\begin{align*}
    \lmp{c}\colon& M\to M, m\mapsto cm\text{ and}\\
    \rmp{d}\colon& M\to M, m\mapsto md.
\end{align*}
This yields the chain complex and, if $M$ is a finite-dimensional $\ftwo$-vector space equipped with a basis, a quantum code
\begin{equation}
\label{eq:complexofbimodulecodes}
\begin{tikzcd}[column sep= 40pt]
   \Code(M,c,d)\colon M & {M_{\ver} \oplus M_{\hor}} & M.
    \arrow["{(\lmp c, -\rmp d)}"', from=1-2, to=1-1]
    \arrow["\left(\begin{smallmatrix} \rmp d \\ \lmp c \end{smallmatrix}\right)"', from=1-3, to=1-2]
\end{tikzcd}
\end{equation}
\begin{remark}
\begin{enumerate}
\item For $R=\ftwo[G]$ and $M=R$, these yield the two-block group algebra codes from \cite{linQuantumTwoblockGroup2023}.
\item If $M=R$ and $R$ is commutative, $\Code(R,c,d)=\Code(c,d)$ and one recovers the construction described in \Cref{sec:quantumcodesfromcommutativerings}.

\item The interested reader may generalize the fundamental sequence, see \Cref{eq:fundamentalsequence}, to apply to the first homology of $\Code(M,c,d).$
\end{enumerate}
\end{remark}

\end{document}